\documentclass[12pt,preprint]{aastex631}

\usepackage{color}
\usepackage{mathrsfs}
\usepackage{amsmath}
\def\VEC#1{\mbox{\boldmath $#1$}}

\shorttitle{1D simulation of Alfv\'{e}n wave in black cylinder spacetime}
\shortauthors{Koide et al.}

\begin{document}


\title{One-dimensional force-free numerical simulations of Alfv\'{e}n waves
around a spinning black string}

\vspace{1cm}


\author{Shinji Koide}
\affil{Department of Physics, Kumamoto University, 
2-39-1, Kurokami, Kumamoto, 860-8555, JAPAN}
\email{koidesin@kumamoto-u.ac.jp}
\author{Sousuke Noda}
% \affil{Miyakonojo College, National Institute of Technology (KOSEN),
% 473-1, Yoshio-cho, Miyakonojo-shi, Miyazaki 885-8567, Japan}
%% \affil{Center for Gravitation and Cosmology, College of Physical Science and Technology,
%% Yangzhou University, Yangzhou 225009, China}
\affil{National Institute of Technology, Miyakonojo College, Miyakonojo 885-8567, Japan}
\affil{Center for Gravitation and Cosmology, College of Physical Science and Technology, 
Yangzhou University, Yangzhou 225009, China}
\affil{School of Aeronautics and Astronautics, Shanghai Jiao Tong University, 200240 Shanghai, China}
\author{Masaaki Takahashi}
\affil{Department of Physics and Astronomy, Aichi University of Education,
Kariya, Aichi 448-8542, Japan}
%% \author{Yasusada Nambu, Takuma Tsukamoto}
\author{Yasusada Nambu}
\affil{Department of Physics, Graduate School of Science, Nagoya University,
Chikusa, Nagoya 464-8602, Japan}



\begin{abstract}
We performed one-dimensional force-free magnetodynamic numerical simulations of 
the propagation of Alfv\'{e}n waves
along magnetic field lines around a spinning black-hole-like object,
the Banados--Teitelboim--Zanelli black string,
to investigate the dynamic process of wave propagation and 
energy transport with Alfv\'{e}n waves.
% We used the force-free linear equation derived by \citet{noda20}.
We considered axisymmetric and stationary magnetosphere
and perturbed the background 
magnetosphere to obtain the linear wave equation for the Alfv\'{e}n wave mode.
% derived by \citet{noda20}, which comes from the leading order of the perturbation.
The numerical results show that the energy of Alfv\'{e}n waves monotonically
increases as the waves propagate outwardly along the rotating curved magnetic 
field line around the ergosphere,
where energy seems not to be conserved,
in the case of energy extraction from the black string 
by the Blandford--Znajek mechanism.
% The energy of the Alfv\'{e}n wave increases by the work of the torque from the
% background magnetic-field line. Inversely, 
% the reaction of the torque induces the azimuthally oscillating wave,
% which is identified by the fast mode.
% The total energy of the Alfv\'{e}n wave and the fast
% wave is conserved. The fast wave never influences the
% Alfv\'{e}n wave linearly because the amplitude of the induced wave is the
% second order of the perturbation of the Alfv\'{e}n wave.
The apparent breakdown of energy conservation suggests the existence of
an additional wave induced by the Alfv\'{e}n wave.
Considering the additional fast magnetosonic wave
induced by the Alfv\'{e}n wave,
the energy conservation is recovered.
Similar relativistic phenomena, such as the amplification of
Alfv\'{e}n waves and induction of fast magnetosonic waves,
are expected around a spinning black hole.
% The turbulence in the ergosphere around the hole may cause the Alfv\'{e}n wave.
\end{abstract}



\section{Introduction} \label{sec1}

\begin{figure} % [H]
\begin{center}
\includegraphics[width=9cm]{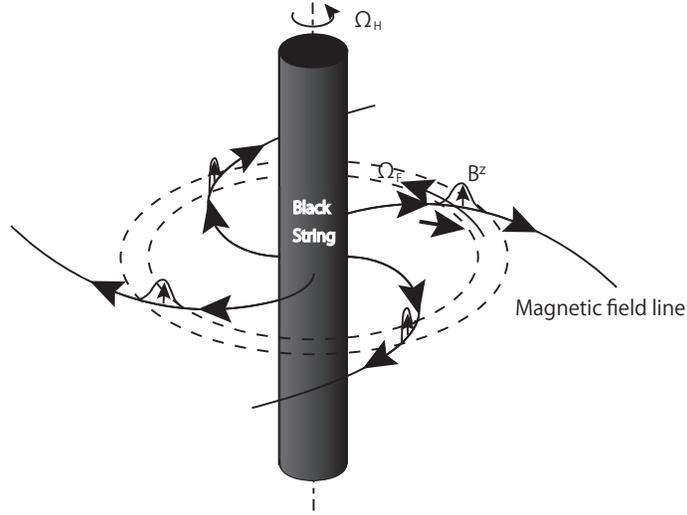}% Here is how to import EPS art
\end{center}
\caption{Schematic of Alfv\'{e}n wave propagation along 
field lines of a force-free magnetic field
around a spinning black string.
%  (a) Face on. (b) Oblique. 
\label{pontie_zentai2}}
\end{figure}

High-energy phenomena, such as
relativistic jets from active galactic nuclei and in gamma-ray bursts, are 
powered by the rotational energy of
a spinning black hole as suggested by general relativistic
magnetohydrodynamic (GRMHD) numerical simulations \citep{koide03,
komissarov05,koide06,gammie03,
mckinney06,mckinney09,eht19,porth19,
mizuno04,nagataki09,paschalidis15,ruiz16,kathirgamaraju19}. 
The GRMHD simulations have shown that the rotational energy of the black hole
is extracted through a magnetic field, and the energy is transported outwardly as
the propagating twist of the magnetic field lines. 
\citet{mizuno04} and \citet{koide06} reported the propagation of the field line
twist as torsional Alfv\'{e}n waves.
Torsional Alfv\'{e}n waves have also been investigated in solar
coronal heating through analytical and numerical methods \citep{musielak07,antolin08}. 
In the studies on the solar corona, the magnetic field lines
of torsional Alfv\'{e}n waves are helical, transverse, and
incompressive. Therefore, the wave is a shear Alfv\'{e}n wave or an Alfv\'{e}n wave.
On the other hand, torsional Alfv\'{e}n waves with spiral magnetic field lines
have not been investigated in solar physics
because such spiral magnetic field lines are not related to coronal heating,
solar flare, and other related phenomena in solar physics. 
Spiral torsional Alfv\'{e}n waves were shown in the GRMHD
numerical simulations for relativistic outflow formation with
an initially radial magnetic field around a rapidly
spinning black hole \citep{koide04,komissarov04}. Such spiral torsional Alfv\'{e}n waves
are compressive, and they are identified by fast magnetosonic waves 
(fast waves). When we consider the perturbative displacement of magnetic fields
perpendicular to axisymmetric magnetic surfaces, the wave
is incompressive, i.e., an Alfv\'{e}n wave.
%
% Besides the relativistic AGN jets,
% Alfv\'{e}n wave plays a significant role in the process of the
% high energy astrophysical phenomena associated with a black hole.
% Thus, it is important to clarify the basic physics of Alfv\'{e}n wave around
% a black hole. 
%
Under the force-free condition,
\citet{noda20} derived a linear equation of an Alfv\'{e}n wave
propagating along a magnetic field line around 
a Banados--Teitelboim--Zanelli (BTZ) black string.
The BTZ spacetime is analogous to the spacetime at the equatorial plane 
of a spinning black hole, but its
translation symmetry is along the rotating axis, and the horizon is cylindrical
as shown in Fig. \ref{pontie_zentai2} . 
% To derive the linear equation, they assumed the force-free condition for
% electromagnetic field.
Considering the perturbation of an electromagnetic field,
a stationary solution is used as a background equilibrium 
around the central object.
To obtain an equilibrium solution around a black hole, it is necessary
to solve the general relativistic Grad--Shafranov (GS) equation \citep{blandford77}.
For a field around a spinning black hole, 
there is no analytic global solution of the GS equation.
For the BTZ black string, there is a global analytic solution
of the equilibrium state; thus, we discuss the essence of phenomena in spacetime
around a spinning black hole.
Using the stationary scattering analytical method of the
linear force-free equation, \citet{noda20} showed that
Alfv\'{e}n waves are reflected around static limit surfaces with a reflection
rate greater than unity, indicating superradiance. 
% \citet{noda20} showed the sinusoidal solution of the linear equation
% with respect to the torsional Alfv\'{e}n wave and proposed
% the superradiance due to the wave around the black string
% using the eikonal approximation.
%%% As far as we use the stationary scattering method, we cannot catch 
The dynamic causal process of the waves cannot be clearly observed using
stationary scattering methods. In this paper, 
we perform numerical simulations of the dynamic processes
of Alfv\'{e}n waves to investigate the details of energy transport
due to the waves. We found an unexpected relativistic phenomenon
of a wave that was induced by Alfv\'{e}n waves.
% The relation between superradiance and the 
% Blandford--Znajek mechanism \citep{blandford77} is also discussed by \citet{noda20}.
% Our original motivation of this study was the verification of the remarks
% in \citet{noda20} by numerical calculations.

To investigate the dynamic processes of the Alfv\'{e}n wave
and energy transport due to the wave,
we performed force-free magnetodynamic (FFMD) simulations of
Alfv\'{e}n waves propagating along the rotating curved magnetic field lines
around a spinning BTZ black string using the linear equation derived by \citet{noda20}.
The numerical results show that the energy of Alfv\'{e}n waves increases
monotonically as the wave propagates outwardly along a rotating curved
magnetic field line, where the black string rotational energy
is extracted by the Blandford--Znajek mechanism, especially around the ergosphere. 
The amplification of the Alfv\'{e}n wave is based on the work of the torque
from the background magnetic-field line.
On the other hand, 
the reaction of the torque induces the azimuthally oscillating wave 
along the background magnetic-field line. 
The induced wave is identified by the fast wave.
The total energy of the induced second-order fast wave and
first-order Alfv\'{e}n wave is conserved. 
The amplitude of the induced fast wave is 
the second order of the perturbation of the Alfv\'{e}n wave, 
and then, the fast wave, higher-order perturbation, does not influence the Alfv\'{e}n wave.

In Section \ref{secalf}, we explain the theoretical background of
Alfv\'{e}n waves along the stationary background magnetic-field lines around a
black string.
In Section \ref{sec3}, we investigate the conservation law of energy and angular momentum 
due to the Alfv\'{e}n wave and show an apparent breakdown of the force-free conditions
with respect to the azimuthal component of the Lorentz force
in the second-order perturbation and the energy conservation law 
of Alfv\'{e}n waves. To recover the force-free condition,
we observed that the second-order variable $\chi$ should be introduced, which expresses
the fast wave. 
% In this section, we use the corotating natural coordinates, which makes
% the calculations simple.
%% we derive the time evolution equation of $\chi$ and confirm the recovery 
%% of the conservation law of the angular momentum
%% in the natural coordinates, which simplify the calculations of
%% the electromagnetic fields of the waves. 
% In Section \ref{sec4}, 
We further show the 
recovered conservation law of energy in the original coordinates
of the BTZ spacetime with $\chi$.
In Section \ref{secmet}, we summarize the numerical methods.
The numerical results are shown in Section \ref{secres}.
We summarize and discuss the results in Section \ref{secdis}.
In this paper, 
we use the natural unit system, where the light speed and gravitation constant
are unity (i.e., $c=1$ and $G=1$).

\newpage

\section{Background magnetosphere and Alfv\'{e}n wave
\label{secalf}}

We investigate force-free Alfv\'{e}n wave along background magnetic field
around a spinning black string,
%%% In any reference frame $x^\mu$, we write the line element $ds$ with
%%% the metric $g_{\mu\nu}$ as
%%% \begin{equation}
%%% ds^2 = g_{\mu\nu} dx^\mu dx^\nu 
%%% = - \alpha^2 dt^2 + \gamma_{ij} (dx^i + \beta^i dt) (dx^j + \beta^j dt) 
%%% \end{equation}
of which line element is given by the following metric of the BTZ black string spacetime 
\begin{equation}
ds^2 = g_{\mu\nu} dx^\mu dx^\nu 
= - \alpha^2 dt^2 + \gamma_{ij} (dx^i + \beta^i dt) (dx^j + \beta^j dt) 
= - \alpha^2 dt^2 + \frac{1}{\alpha^2} dr^2
+ r^2 (d \varphi - \Omega dt)^2 + dz^2 ,
\end{equation}
where
Greek indices such as $\mu$ and $\nu$ run from 0 to 3,
while Roman indices such as $i$ and $j$ run from 1 to 3,
$\displaystyle \alpha^2 = \frac{1}{r^2 \ell^2} \left [ r^4 - Mr^2 + \left ( \frac{Ma}{2} \right )^2 \right ]$, 
$\displaystyle \Omega = \frac{M a}{2 r^2 \ell}$, 
$\displaystyle \gamma = \det (\gamma_{ij}) = \left ( \frac{r}{\alpha} \right )^2$,
and $\displaystyle -g = \alpha^2 \gamma = r^2$ \citep{jacobsen19}.
%% To distinguish the lapse function of the BTZ spacetime and the natural coordinates,
%% we denote the lapse function of BTZ spacetime by $\underline{\alpha}^2$.
$M$ is the mass line density of the black string 
and $a$ is the spin parameter of the black string,
whose line density of angular momentum is given as $J=a M \ell$.
$\ell$ denotes the anti-deSitter curvature scale related to the negative cosmological constant as 
$\Lambda_3 = - 1/\ell^2$.
Through this paper, we set $\ell$ unity.
The outer horizon of this spacetime is located at $r=r_{\rm H} = [(1+\sqrt{1-a^2})M/2]^{1/2}$
where $\alpha(r_{\rm H})=0$. Each constant-$z$ slice of the spacetime
is the (2+1)-dimension
%% Banados-Teitelboim-Zanelli (
BTZ black string
%% ) black string 
spacetime \citep{banados92}, and hence,
the horizon geometry is cylindrical.
The angular velocity at the horizon is given by
\begin{equation}
\Omega_{\rm H} \equiv \Omega(r_{\rm H}) = \frac{1}{\ell}
\left ( \frac{a}{1 + \sqrt{1-a^2}} \right ).
\end{equation}
%%% Hereafter, for simplicity, we set $\ell$ unity.
%%% Then, we have the metric $g_{\mu\nu}$ as
%%% \begin{equation}
%%% g_{\mu\nu} = \left ( \begin{array}{cccc}
%%% - \alpha^2 + r^2 \Omega^2 & 0 & -r^2 \Omega & 0 \\
%%% 0 & \displaystyle \frac{1}{\alpha^2} & 0 & 0 \\
%%% -r^2 \Omega & 0 & r^2 & 0 \\
%%% 0 & 0 & 0 & 1 \end{array} \right ),
%%% \displaystyle g^{\mu\nu} = \left ( \begin{array}{cccc}
%%% \displaystyle - \frac{1}{\alpha^2} & 0 & \displaystyle -\frac{\Omega}{\alpha^2} & 0 \\
%%% 0 & \alpha^2 & 0 & 0 \\
%%% \displaystyle - \frac{\Omega}{\alpha^2} & 0 & \displaystyle \frac{1}{r^2} - \frac{\Omega}{\alpha^2} & 0 \\
%%% 0 & 0 & 0 & 1 \end{array} \right ).
%%% \end{equation}

% \paragraph{Euler potential の導入}
%% To express the force-free electromagnetic field,
%% we introduce 

\subsection{Background magnetosphere}
When the electromagnetic field satisfies the force-free condition
$J^\mu F_{\mu\nu}=0$, 
the inhomogeneous Maxwell equation $\nabla_\mu F^{\mu\nu}=-J^\nu$ 
yields the  description of the field by the
Euler potentials $\phi_1$ and $\phi_2$ as:
% ※$\epsilon^{\mu\nu\rho\sigma}$の符号はNoda et al.(2019)と逆になっている。
\footnote{The sign of $\epsilon^{\mu\nu\rho\sigma}$ is opposite
from that of \citet{noda20}}
\begin{eqnarray}
F_{\mu \nu}  &=& \partial_\mu \phi_1 \partial_\nu \phi_2 
- \partial_\nu \phi_1 \partial_\mu \phi_2 ,\\
{}^* F^{\mu \nu} &=& \frac{1}{2} \epsilon^{\mu\nu\rho\sigma} F_{\rho\sigma},
\end{eqnarray}
where $\epsilon^{\mu\nu\rho\sigma}$ is the completely antisymmetric tensor
and is determined by $\epsilon^{0123}=1/\sqrt{-g}$ \citep{uchida97a,uchida97b}.
When we use Euler potential, one of the (homogeneous) Maxwell equation, 
$\nabla_\mu {}^* F^{\mu\nu}=0$ 
is satisfied identically. 
The (inhomogeneous) Maxwell equation $\nabla_\lambda F^{\lambda \mu} = - J^\mu$ 
and force-free field $J^\mu F_{\mu\nu} =0$ yield
equations for $\phi_1$ and $\phi_2$,
%% \begin{equation}
%% \partial_\lambda \phi_i \partial_\nu [
%% \sqrt{-g} (g^{\lambda \alpha} g^{\nu \beta} - g^{\nu \alpha} g^{\lambda \beta})
%% \partial_\alpha \phi_1 \partial_\beta \phi_2 ] = 0 , \verb!   !(i=1,2)
%% \label{nodaeq04}
%% \end{equation}
%% When we use 
%% we write Eq. (\ref{nodaeq04}) by
\begin{equation}
\partial_\lambda \phi_i \partial_\nu [
\sqrt{-g} W^{\lambda\alpha\nu\beta}
\partial_\alpha \phi_1 \partial_\beta \phi_2 ] = 0  \verb!   !(i=1,2),
\label{nodaeq04w}
\end{equation}
where $W^{\lambda\alpha\nu\beta} = g^{\lambda\alpha} g^{\nu\beta}
- g^{\lambda\beta} g^{\alpha\nu}$.

% \paragraph{Normal observer frameと電磁場の定義}

The 4-velocity of the normal observer frame $x^{\tilde{\mu}}$ is given by
\begin{equation}
N^\lambda = \left (\frac{1}{\alpha}, - \frac{\beta^i}{\alpha} \right ),
N_\lambda = (- \alpha, 0, 0, 0),
\end{equation}
where $\beta_i = g_{0i}$, $\beta^i = g^{ij} \beta_j$.
The electromagnetic field is defined by
%% \begin{eqnarray}
%% E^\lambda &=& F^{\lambda \nu} N_\nu = E^{\tilde{\lambda}}, \\
%% B^\lambda &=& {}^* F^{\lambda \nu} N_\nu = B^{\tilde{\lambda}}.
%% \end{eqnarray}
%% Then, we have
\begin{eqnarray}
E^\lambda &=& F^{\lambda \nu} N_\nu =  - \alpha F^{\lambda 0} 
= - \alpha (\partial^\lambda \phi_1 \partial^0 \phi_2 
- \partial^0 \phi_1 \partial^\lambda \phi_2 ), \\
B^\lambda &=& {}^* F^{\lambda \nu} N_\nu =  - \alpha {}^* F^{\lambda 0} = 
\frac{1}{\sqrt{\gamma}} \eta^{0 \lambda \alpha \beta}
\partial_\alpha \phi_1 \partial_\beta \phi_2 .
\end{eqnarray}

% \begin{section} BTZ spacetime

%% We use the metric of the BTZ space time 
% In $x^\mu = (t, r, \varphi, z)$ coordinates,
%% to express the spacetime around the spinning black string.

% \paragraph{平衡解}
According to Eq. (\ref{nodaeq04w}),
the steady state solution of the force-free field around the black string
is given by \citet{jacobsen19} as
\begin{equation}
\bar{\phi}_1 = - \int g_{zz} dz = - z, \verb!   !
\bar{\phi}_2 = 
%%% h(r) + \varphi - \Omega_{\rm F} t = 
\frac{I}{2 \pi} \int \frac{dr}{r \alpha^2}+ \varphi - \Omega_{\rm F} t,
\label{eqequisol}
\end{equation}
where 
%%% $\displaystyle h'(r) = \frac{I}{2 \pi} \frac{1}{r \alpha^2}$ and
$I$ is a constant of total current and $\Omega_{\rm F}$ is a constant
of the angular velocity of the magnetic field line. 
Note that $\phi_1$ corresponds to the stream function and contour of $\bar{\phi}_1$ 
expresses the steady magnetic surfaces.
In the present case, $z=$ const. presents the steady magnetic surface.
We have
\begin{eqnarray}
\bar{E}^z &=& - \frac{1}{\alpha} (\Omega - \Omega_{\rm F}), \verb!  !
\bar{B}^r = \frac{1}{\sqrt{\gamma}} = \frac{\alpha}{r}, \verb!  !
\bar{B}^\varphi = - \frac{I}{2 \pi r^2 \alpha} = - \frac{1}{\sqrt{\gamma}} 
\frac{I}{2 \pi} \frac{1}{r \alpha^2}, \\
\bar{E}^r &=& \bar{E}^\varphi = 0, \verb!  ! \bar{B}^z = 0 .
\end{eqnarray}
The continuity of the scalar of the field $F^{\lambda\kappa}F_{\lambda\kappa}$
at the horizon
yields the condition with respect the total current,
\begin{equation}
I = 2 \pi r_{\rm H} (\Omega_{\rm H} -\Omega_{\rm F}).
\label{cond4i}
\end{equation}
The corotating vector of the field line, 
$\xi_{(\rm F)}^\nu \equiv \xi_{(t)}^\nu + \Omega_{\rm F} \xi_{(\varphi)}^\nu$ is
parallel to the 4-velocity of corotating observer with  the magnetic field line.
%%% The 4-velocity of rotation of magnetic field line is given by
%%% $\displaystyle u^\mu_{\rm F} = \frac{1}{\sqrt{-\Gamma}} \xi^\mu_{\rm (F)}$,
%%% where $\Gamma$ is 
The norm of $\xi_{(\rm F)}^\nu$ is
\begin{equation}
\Gamma = g_{\mu\nu} \xi^\mu_{\rm (F)} \xi^\nu_{\rm (F)}
=- \alpha^2 + r^2 (\Omega - \Omega_{\rm F})^2 = - \gamma_{\rm F} (r^2 -r_{\rm LS}^2),
\end{equation}
with $\gamma_{\rm F}=1 - \Omega_{\rm F}^2$.
%%% \begin{equation}
%%% \Gamma = g_{\lambda \nu} \xi_{\rm F}^\lambda \xi_{\rm F}^\nu
%%% = - \alpha^2 + r^2 (\Omega - \Omega_{\rm F})^2.
%%% \end{equation}
The real root of $\Gamma=0$ gives the location of the light surface uniquely,
which is the causal boundary for Alfv\'{e}n waves and we denote its location
by $r=r_{\rm LS}=\sqrt{(1-a \Omega_{\rm F})/(1-\Omega_{\rm F}^2)}$.

% \section{Magnetically Natural frame}

%% We first rewrite Eq. (\ref{nodaeq10}) in terms of a parameter along a magnetic field
%% line $\sigma$ and the time coordinate for a corotating observer of the magnetic field
%% line $\tau$. The coordinates $(\tau,\sigma,\rho)$ are introduced through the
%% following frame transformation:
To reduce the calculations to a simple form, following \citet{noda20},
we utilize the coordinates $(T, X, \rho, z)$ defined as
\begin{align}
T &= t + \int I_0 dX , & \displaystyle t &= T - \int I_0 dX ,\\
X &= r , & r &= X, \\
\rho &= \varphi - \int I_1 dr - \Omega_{\rm F} t  , & \displaystyle \varphi &= \rho + \int I_2 dX + \Omega_{\rm F} T  ,\\
  z &= z , & z &= z
\end{align}
where $\displaystyle I_0 = I' X \frac{\Omega - \Omega_{\rm F}}{{\alpha}^2 \Gamma}$,
$\displaystyle I'=\frac{I}{2 \pi}$,
$\displaystyle I_2 = \frac{I'}{X {\alpha}^2 \Gamma} ({\alpha}^2 - X^2 \Omega (\Omega - \Omega_{\rm F}))$,
$\displaystyle I_1 = I_2 + \Omega_{\rm F} I_0 = - \frac{I'}{{\alpha}^2 X}$.
%% $\rho$ is $\phi_2$ itself, and each $\rho = \rm const$ gives a 
%% magnetic field line. Therefore, $\rho$ is a coordinate perpendicular field line.
%
In this coordinate system,
$X$ is the coordinate along the magnetic field line, then we call this
coordinates system $(T, X, \rho, z)$ ``corotating magnetically natural frame" or
simply ``corotating natural frame".
In this section, hereafter, we use the corotating natural coordinates.
The metric tensor of the corotating natural frame is as follows:
\begin{equation}
g_{\mu\nu} = \left ( \begin{array}{cccc}
\Gamma & 0 & -X^2 W & 0 \\
0 & Y & J & 0 \\
-X^2 W & J & X^2 & 0 \\
0 & 0 & 0 & 1 \end{array} \right ), \verb!    !
\displaystyle g^{\mu\nu} = \left ( \begin{array}{cccc}
\displaystyle - \frac{\gamma_{\rm N}}{X^2} & J W & \displaystyle - Y W & 0 \\
J W & {\alpha}^2 & \displaystyle \frac{I'}{X} & 0 \\
\displaystyle - Y W & \displaystyle \frac{I'}{X} & \displaystyle - \frac{\Gamma Y}{X^2} & 0 \\
0 & 0 & 0 & 1 \end{array} \right ),
\end{equation}
where $\displaystyle W = \Omega - \Omega_{\rm F}$, 
$\displaystyle Y = \frac{1}{{\alpha}^2 \Gamma}(\Gamma - I'^2)$,
$\displaystyle J = \frac{I'X}{\Gamma}$, and
$\displaystyle 
\gamma_{\rm N} \equiv {\rm det} (g_{ij}) 
= \frac{X^2}{{\alpha}^2 \Gamma^2} (\Gamma^2 - I'^2 X^2 W^2)$. 
We also have $\sqrt{-g} = X$ and the lapse function, $\displaystyle \alpha^2_{\rm N}
= \frac{X}{\gamma_{\rm N}}$ in the coordinates.
In the corotating natural coordinates, the steady-state force-free solution
given by Eq. (\ref{eqequisol}) is described simply as
\begin{equation}
\bar{\phi}_1 = -z, \verb!   !
\bar{\phi}_2 = \rho.
\label{eqequsolconaf}
\end{equation}

%% To eliminate the cross term of $\tau$ and $\sigma$, we choose another
%% set of coordinates $(T, X)$ on the field sheet, defined as
%% \begin{equation}
%% \tau = - I \int dX X \frac{\Omega - \Omega_{\rm F}}{\alpha^2 \Gamma} + T,
%% \hspace{0.5cm} \sigma = X,
%% \end{equation}
%% where we have $\partial_T = \partial_\tau$ and
%% $\partial_X = \partial_\sigma - I \sigma (\Omega - \Omega_{\rm F})/(\alpha^2 \Gamma) 
%% \partial_\tau$. 

\subsection{Wave equation for Alfv\'{e}n wave}

% \paragraph{Perturbation of Alfv\'{e}n wave}

%%% We consider the perturbation of Alfv\'{e}n wave on the steady state
%%% force-free electromagnetic field around a spinning black cylinder:
%%% $\phi_1 = \bar{\phi}_1 + \delta \phi_1 = - z \delta \phi_1$, $\phi_2 = \bar{\phi}_2$
%% Taking the first-order terms of Eq. (\ref{nodaeq04}), we obtain the following
%% equations for the perturbations $\delta \phi_1 = \phi_1 - \bar{\phi}_1$ 
%% and $\delta \phi_2= \phi_2 - \bar{\phi}_2$:
%% \begin{eqnarray}
%% \partial_\nu \phi_2 \partial_\lambda (
%% \sqrt{-g} \partial^{[\lambda} \delta \phi_1 \partial^{\nu ]}\phi_2) &=& 0,
%% \label{nodaeq10} \\
%% \partial_j (\sqrt{-g} \partial^j \delta \phi_2) &=& 0,
%% \label{nodaeq11}
%% \end{eqnarray}
%% where $j=t,r,\varphi$ and the square bracket represents the anticommutator.
%% The perturbation $\delta \phi_2$ corresponds to fast magnetosonic wave, whereas
%% the perturbation $\delta \phi_1$ represents the Alfv\'{e}n wave, which
%% propagates along a magnetic field line on a magnetic surface.

\begin{figure}%[p]
\begin{center}
\includegraphics[scale=1.3]{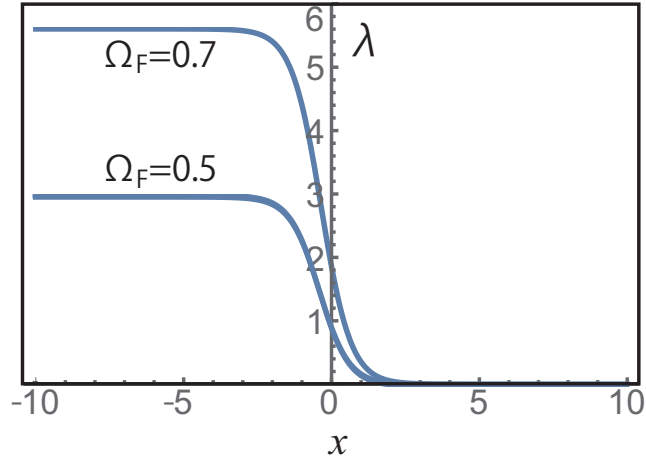}% Here is how to import EPS art
\caption{Profiles of $\lambda$ in the cases of
$a=0.9$ and $\Omega_{\rm F}=0.5$ ($0 < \Omega_{\rm F} < \Omega_{\rm H}$)  and
$a=0.9$ and $\Omega_{\rm F}=0.7$ ($\Omega_{\rm F} > \Omega_{\rm H}$).
%%  In this case, $\kappa$ vanishes.
\label{prolambda}}
\end{center}
\end{figure}

In the background magnetosphere given by Eq. (\ref{eqequsolconaf}), 
we consider the propagation of the force-free waves.
The infinitesimally small perturbation to the Euler potential 
$\phi_i \longrightarrow \phi_i + \delta \phi_i$
is given by the displacement vector $\zeta^\lambda$ as
$\delta \phi_i = \zeta^\lambda \partial_\lambda \phi_i$.
We focus on the wave propagating on the magnetic surface given by $z=0$
and oscillating perpendicularly to the magnetic surface:
$\zeta^\lambda =(0, 0, 0, \zeta^z)$. Then,
the perturbed Euler potentials are
\begin{equation}
\phi_1 = \bar{\phi}_1 + \delta \phi_1 = - z + \delta \phi_1, \verb!   ! 
\phi_2 = \bar{\phi}_2 = \rho,
\end{equation}
because $\delta \phi_2 = \zeta^2 \partial_z \phi_2 = \zeta^z \partial_z \rho = 0$.
%%% where the magnetic field $\phi_1 = \rm const.$ shifts toward
%%% $z$-direction, for example, the steady magnetic surface $z=0$ shifts to the
%%% surface $z=\delta \phi_1$
We assume the perturbation is translationary invariant for $z$-direction,
\begin{eqnarray}
\delta \phi_1 &=& \psi (T, X, \rho) .
\end{eqnarray}
%
%% The differential operators with respect to the new coordinates are
%% $\partial_\tau = \partial_t + \Omega_{\rm F} \partial_\varphi$ and 
%% $\partial_\sigma = \partial_r - I/(r \alpha^2) \partial_\varphi$.
%% In these coordinates, Eq. (\ref{nodaeq10}) yields
%% \begin{equation}
%% - \left ( 1 + \frac{I^2}{\alpha^2} \right ) \partial^2_\tau \delta \phi
%% - \alpha^2 \sigma  \partial_\sigma \left [ \frac{\Gamma}{\sigma} \left (
%% \partial_\sigma - \frac{I \sigma (\Omega - \Omega_{\rm F})}{\Gamma \alpha^2} \partial_\tau
%% \right ) \delta \phi \right ]
%% + I \sigma (\Omega - \Omega_{\rm F}) \partial_\sigma \partial_\tau \delta \phi
%% - \mu^2 \sigma^2 \alpha^2 C_2 \delta \phi = 0,
%% \label{nodaeq13}
%% \end{equation}
%% where $C_2 \equiv I^2/(\sigma^2 \alpha^2) - (\Omega - \Omega_{\rm F})^2/\alpha^2
%% + 1/\sigma^2$ and 
%%% $\xi_{\rm F}^\nu = (\partial_t)^\nu + \Omega_{\rm F} (\partial_\varphi)^\nu$:
%
In the corotating coordinates, linearization of Eq. (\ref{nodaeq04w}) with $i=2$ yields
\begin{equation}
%%% \partial_\lambda (\sqrt{-g} W^{\lambda\alpha\mu\beta} &&
%%% \partial_\alpha \psi \partial_\beta \bar{\phi}_2 ) \partial_\mu \bar{\phi}_2 = 0, \\
\partial_\lambda (\sqrt{-g} W^{\lambda\alpha\mu\beta}
\partial_\alpha \psi \partial_\beta \bar{\phi}_2 ) \partial_\mu \bar{\phi}_1
+ \partial_\lambda (\sqrt{-g} W^{\lambda\alpha\mu\beta} 
\partial_\alpha \bar{\phi}_1 \partial_\beta \bar{\phi}_2 ) \partial_\mu \psi = 0,
\label{eq4i2} 
\end{equation}
%%% respectively.
%%% When we take $i=1$ in Eq. (\ref{nodaeq04w}), we have a trivial equation
%%% because $W^{\lambda\alpha z\rho}=0$ ($\alpha, \lambda \neq z$).
and can be simplified as
%%%% Eq. (\ref{eq4i2}) yields
%% \begin{equation}
%% - \left ( 1 - \frac{I'^2}{\Gamma} \right ) \partial^2_T \delta \phi
%% + \alpha^2 \gamma_{\rm F} \left ( X + \frac{X^2_{\rm LS}}{X} \right ) \partial_X \delta \phi
%% - \alpha^2 \Gamma \partial^2_X \delta \phi
%% - \mu^2 \sigma^2 \alpha^2 C_2 \delta \phi = 0,
%% \label{nodaeq16d}
%% \end{equation}
\begin{equation}
 \left ( 1 - \frac{I'^2}{\Gamma} \right ) \partial^2_T \psi
+ {\alpha}^2 X \partial_X \left ( \frac{\Gamma}{X}  \partial_X \psi \right ) = 0.
%% + \mu^2 \sigma^2 \alpha^2 C_2 \psi = 0.
\label{nodaeq16dd}
\end{equation}
%%% where $\gamma_{\rm F} \equiv 1 - \Omega_{\rm F}^2$, $X_{\rm LS} = r_{\rm LS}
%%% =\sqrt{(1-a \Omega_{\rm F}) M/\gamma_{\rm F}}$.
Eq. (\ref{nodaeq16dd}) is identified to the time evolution version of Eq. (16) 
in \citet{noda20}.
%% Then, we call Eq. (\ref{nodaeq16dd}) ``Noda equation" in this paper.
%%% We have $\Gamma = - \gamma_{\rm F} (X - X_{\rm LS}^2)$. 
Note that Eq. (\ref{nodaeq16dd}) does not have a derivative term with respect
to $\rho$. This means the perturbation $\delta \phi_1$ propagates independently
on a two-dimensional sheet spanned by $T$ and $X$, called a field sheet,
which represents the time evolution of a magnetic field line.
Therefore, we can identify $\psi$ as the perturbation of an Alfv\'{e}n wave.
Furthermore, we can assume the perturbation of the field is axisymmetric,
without loss of generality, $\psi (T, X, \rho)=\psi(T,X)$,
i.e. $\psi$ is axisymmetric.
%%% because Eq. (\ref{nodaeq16dd}) is not partial differential equation with respect
%%% to the variable $\rho$ and we have a solution $\psi (x, \rho, T)=\psi(x,T)$.
Introducing the dimensionless ``tortoise" coordinate $x$ as
\begin{equation}
\frac{dx}{dX} = \frac{X}{X^2-X_{\rm LS}^2} 
= \frac{- \gamma_{\rm F} X}{\Gamma}
\hspace{0.5cm} (-\infty < x < \infty),
\end{equation}
where $X_{\rm LS} = r_{\rm LS}$,
we have $\displaystyle x = \frac{1}{2} \log (X^2 - X^2_{\rm LS})$ and Eq. (\ref{nodaeq16dd}) reads
\begin{equation}
- \left [ 1 - \frac{I'^2}{\Gamma} \right ] \frac{\partial^2 \psi}{\partial T^2}
+ {\alpha}^2 \gamma_{\rm F} 
\frac{X^2}{X^2-X_{\rm LS}^2} \frac{\partial^2 \psi}{\partial x^2} = 0.
%% - \mu^2 X^2 \alpha^2 C_2 \psi = 0
\label{nodaeq19d}
\end{equation}
If we use the variable
\begin{eqnarray}
\lambda(x) &=& \frac {X^2-X_{\rm LS}^2 + {I'}^2/\gamma_{\rm F}}
{\gamma_{\rm F} {\alpha}^2 X^2}
= \frac{I'^2 - \Gamma}{\gamma_{\rm F}^2 \alpha^2 X^2}
= \frac{I'^2 + \alpha^2 - X^2 (\Omega - \Omega_{\rm F})^2}{\gamma_{\rm F}^2 \alpha^2 X^2} , 
%% \kappa(x) &=& \frac{\mu^2 (X^2 -X_{\rm LS}^2) C_2 }{\gamma_{\rm F}},
\end{eqnarray}
Eq. (\ref{nodaeq19d}) is written as
\begin{equation}
\lambda \frac{\partial^2 }{\partial T^2} \psi
=  \frac{\partial^2}{\partial x^2} \psi .
%%  - \kappa \delta \phi.
\label{newtontoyeq}
\end{equation}
This equation is identified by the equation of displacement of
the string with non-uniform line density $\lambda(x)$ and
the uniform (unit) tension $T_{\rm str}=1$.
% Then, Eq. (\ref{newtontoyeq}) is called ``simple string-like Noda equation" in this paper.
The profiles of $\lambda$ in the cases of 
$\Omega_{\rm F}=0.5$ and $\Omega_{\rm F}=0.7$ 
with $a=0.9$ ($\Omega_{\rm H} = 0.6268$) 
are shown in Fig. \ref{prolambda}.
It is noted that the rotational energy of the black string is extracted
through the Blandford--Znajek mechanism when $\Omega_{\rm F}$ is larger than zero
and smaller than $\Omega_{\rm H}$. 
%%  and \ref{prolambda2}.
The profile of $\lambda(x)$ indicates the linear Alfv\'{e}n wave, whose wavelength
is much longer than the characteristic length of the gradient of $\lambda$ 
at $x \sim 0$, $l_{\rm c} = 3$, is reflected at the point $x \sim 0$
as remarked by \citet{noda20}.
Only the outwardly propagating Alfv\'{e}n wave can be reflected at $x \sim 0$
like the reflection with a fixed boundary condition,
while the inwardly propagating Alfv\'{e}n wave is not reflected to propagate
smoothly inwardly.
Because $\lambda$ approaches to zero exponentially at $x \longrightarrow \infty$ and 
% when $x \longrightarrow - \infty$ ($r \longrightarrow r_{\rm LS}$)
$\lambda$ approaches to a constant
\begin{equation}
\lambda_{\rm LS} =\left [ 
\frac{I'}{\gamma_{\rm F} {\alpha}(r_{\rm LS}) r_{\rm LS}} \right ]^2
= \left [ \frac{1}{M} \frac{2 r_{\rm H} (\Omega_{\rm F}-\Omega_{\rm H})}
{ a (\Omega_{\rm F}^2 +1) - 2 \Omega_{\rm F} } \right ]^2
\end{equation}
at $x \longrightarrow - \infty$ ($r \longrightarrow r_{\rm LS}$).

%% The value of $\lambda$ of the case with the parameters
%% $\mu^2=0$, $a=0.9$, $\Omega_{\rm F}=0.5$ and $\Omega_{\rm F}=0.7$ is shown in 
%% Figs. \ref{prolambda} and  \ref{prolambda2}, respectively.
%% \begin{figure}%[p]
%% \begin{center}
%% \includegraphics[scale=0.7]{newtonToy2.eps}% Here is how to import EPS art
%% \caption{The profile of $\lambda$ in the case of
%% $\mu^2=0$, $a=0.9$, and $\Omega_{\rm F}=0.7$.
%% In this case, $\kappa$ vanishes.
%% \label{prolambda2}}
%% \end{center}
%% \end{figure}

%% The $z$-component of the magnetic field $B^z$ is given by
%% \begin{eqnarray}
%% B^z = \delta B^z &=& \frac{1}{\sqrt{\gamma}} \frac{\partial \delta \phi}{\partial r} 
%% = \frac{\alpha}{r^2 - r_{\rm LS}^2} \left [
%% \frac{\partial \delta \phi}{\partial x} -
%% \frac{\Omega - \Omega_{\rm F}}{\gamma_{\rm F} \alpha^2} I
%% \frac{\partial \delta \phi}{\partial T} 
%% \right ],
%% \label{calbz}
%% \end{eqnarray}
%% because $\bar{B}^z=0$, where we assume axisymmetry of the field.

\section{Energy and angular momentum conservation 
and force-free condition of the Alfv\'{e}n wave
\label{sec3}}
In this section, we consider the transport of the energy and angular momentum 
with respect to the Alfv\'{e}n wave. 
In the first two subsections ({\S} \ref{subsec1} and {\S} \ref{subsec2}), we use
the corotating natural coordinates $(T, X, \rho, z)$. 
In the last subsection ({\S} \ref{subsec3}), the energy transport in the coordinates
$(t, r, \varphi, z)$ of the BTZ black string is shown.

% \paragraph{Transport of energy angular momentum and conservations laws}

\subsection{Apparent breakdown of energy and angular momentum conservation
\label{subsec1}}

%%% and force-free condition with the solution $\psi$ of Eq. (\ref{newtontoyeq}) only}
%%% The electromagnetic energy and angular momentum are conserved only
%%% in the case of the force-free field. 
%%% To find the order of the perturbation required for investigation
%%% of the force-free condition, 

In this subsection,
we consider only first-order perturbation of the Alfv\'{e}n wave, $\psi$,
and neglect the perturbation of $\phi_2$, where
$\psi$ is given by Eq. (\ref{newtontoyeq}).
The Maxwell equations yield the energy momentum conservation law,
\begin{equation}
\nabla_\nu T^{\mu\nu} = - f_{\rm L}^\mu ,
%%% = - J^\nu {F^\mu}_\nu,
\end{equation}
where $\displaystyle T^{\mu\nu} = {F^{\mu}}_\sigma F^{\nu\sigma} 
- \frac{1}{4} g^{\mu\nu} F_{\lambda\kappa} F^{\lambda\kappa}$ 
is the energy-momentum tensor of electromagnetic field
and $f_{\rm L}^\mu = J^\nu {F^\mu}_\nu$ is the 4-Lorentz force density.
Here, $f_{\rm L}^\mu = 0$ is called the ``force-free condition".

To investigate the energy and momentum transfer due to the Alfv\'{e}n wave,
we have to consider the perturbation of $T^{\mu\nu}$ with the first order or higher 
order.
Hereafter, we note the stationary background (zeroth-order) term, first-order perturbation, 
second-order perturbation, $\cdots$ of a variable $A$ by
$\bar{A}$, $\delta A$, $\delta^2 A$, $\cdots$, i.e., $A = \bar{A} + \delta A + \delta^2 A
+ \cdots$.
%% Eq. (\ref{delsdelmvan}) shows the first order of the energy density 
%% of the Alfv\'{e}n wave vanishes.
When we assume
\begin{equation}
\phi_1 = \bar{\phi}_1 + \delta \phi_1 = -z + \psi, \verb!   !
\phi_2 = \bar{\phi}_2,
\end{equation}
the first order of $T^{\mu\nu}$ is calculated as (see Appendix \ref{appenda})
\begin{equation}
\delta T^\mu_\nu = \delta F^{\mu\lambda} \bar{F}_{\nu\lambda}
+ \bar{F}^{\mu\lambda} \delta F_{\nu\lambda}
- \frac{1}{2} g^\mu_\nu \bar{F}^{\lambda\kappa} \delta F_{\lambda\kappa}.
\label{deltmunu}
\end{equation}
The second and third terms of the right-hand side of Eq. (\ref{deltmunu}) vanish
because of $\partial_z \psi = 0$
% \footnote{
% $\bar{F}^{\mu\lambda} \delta F_{\nu \lambda} 
% = \bar{F}^{\mu z} \delta F_{\nu z} = g^{\mu \rho} \delta F_{\nu z} = 0$.
% $\displaystyle \bar{F}^{\lambda\kappa} \delta F_{\lambda\kappa} =
% \bar{F}_{\lambda\kappa} \delta F^{\lambda\kappa}
% = 2 \bar{F}_{\rho z} \delta F^{\rho z} = 2 W^{\rho\lambda z \rho} \partial_\lambda \psi
% = -2 g^{\rho\rho} g^{\lambda z} \partial_\lambda \psi$=0.}
(see Appendix \ref{appenda}).
The first term of the right-hand side of Eq. (\ref{deltmunu}) is
\begin{equation}
\delta F^{\mu\lambda} \bar{F}_{z\lambda}
= \delta F^{\mu\rho} \bar{F}_{z\rho} = - \delta F^{\mu\rho}
= - W^{\mu\lambda\rho\rho},
\end{equation}
which does not always vanish.
while, if $\mu$ and $\nu$ are not $z$, it vanishes 
% \footnote{$\displaystyle \delta F^{\mu\lambda} \bar{F}_{\nu\lambda}
% = \delta F^{\mu z} \bar{F}_{\nu z} + \delta F^{\mu\rho} \bar{F}_{\nu\rho}
% = \delta F^{\nu z} \delta_{\nu\rho} - \delta^{\mu\rho} \delta_{\nu z}
% = W^{\mu\lambda z\rho} \partial_{\lambda} \psi \delta_{\nu\rho}
% - W^{\mu\lambda \rho\rho} \partial_{\lambda} \psi \delta_{\nu z}
% = - g^{\mu\rho} g^{zz} \partial_z \psi \delta_{\nu\rho}
% - W^{\mu\lambda\rho\rho} \partial_\lambda \psi \delta_{\nu z}
% = W^{\mu\lambda z\rho} \partial_\lambda \psi \bar{F}_{\nu z}=0.$}
(see Appendix \ref{appenda}).
Then, when $\mu$ and $\nu$ are not $z$, we have
\begin{equation}
\delta T^\mu_\nu = 0.
\label{deltzero}
\end{equation}
Eq. (\ref{deltzero}) suggests that as far as we consider the energy and angular momentum
transfer of the Alfv\'{e}n wave along the equatorial plane with respect to the system
with the translational symmetry toward $z$-direction,
$\delta T^{\mu\nu}$ vanishes.
%% It shows that the first-order perturbation of the energy flux density
%% of the Alfv\'{e}n wave vanishes.
This shows the second-order perturbation should be considered
to investigate the energy transport of the linear Alfv\'{e}n wave.
Then, the force-free condition should be satisfied up to second order 
of the perturbation. If the force-free condition of the second-order 
perturbation is broken, the energy and angular momentum of the Alfv\'{e}n wave
is not conserved.
% The conservation of the angular momentum of the Alfv\'{e}n wave is the same as
% that of the Alfv\'{e}n wave energy.
To check the conservation of the energy and the angular momentum, we evaluate
the force-free condition up to the second order of the perturbation
of the Alfv\'{e}n wave.
%
% \paragraph{The check of the force-free condition}
For this purpose, we calculate the Lorentz force,
\begin{equation}
f_\mu^{\rm L} = J^\nu F_{\mu\nu}, \verb!   !
J^\mu = \nabla_\nu F^{\mu\nu} = \frac{1}{\sqrt{-g}} \partial_\nu (\sqrt{-g} F^{\mu\nu}).
\end{equation}
For the background, we can confirm the Lorentz force of the equilibrium vanishes because
the 4-current density of the equilibrium vanishes (see derivation in
Appendix \ref{appenda}), $\bar{f}_\mu^{\rm L} = 0$.
For the first-order perturbation, we confirm the Lorentz force with respect to the first-order perturbation
of Alfv\'{e}n wave vanishes, $\delta f_\mu^{\rm L} = 0$ 
(see the detail derivation in Appendix \ref{appenda}).
It is noted that we have $\delta J^\rho = \delta J^z = 0$
(see Appendix \ref{appenda}).
%
%%% To obtain a guess, first we calculate the Lorentz force of the simplest case:
%%% \begin{equation}
%%% \delta^2 \phi_1 = 0, \verb!   ! \delta \phi_2 =  \delta^2 \phi_2 =0.
%%% \end{equation}

%%%% Last, we check the Lorentz force with respect to the second-order perturbation 
%%%% of Alfv\'{e}n wave.
For the second-order quantities, we have
$\displaystyle 
\delta^2 J^\mu = \frac{1}{\sqrt{-g}} \partial_\nu (\sqrt{-g} \delta^2 F^{\mu\nu})=0$ 
because we neglect the perturbation with respect to $\phi_2$.
Then, we find three components of the  Lorentz force 
except for the $\rho$-component vanish, $\delta^2 f_\mu^{\rm L} = 0$,
because of $\delta J^\rho = 0$ (see Appendix \ref{appenda}).
The $\rho$-component of the Lorenz force is calculated (see also Appendix \ref{appenda}) as
\begin{equation}
\delta^2 f_{\rho}^{\rm L} = \frac{1}{\sqrt{-g}}
\left [ - \partial_X \left \{ \partial_T \psi \left ( \frac{I'}{\Gamma} \partial_T \psi
+ X W \partial_X \psi \right ) \right \}
 + \partial_T \left \{ \partial_X \psi \left (\frac{I'}{\Gamma} \partial_T \psi
+ X W \partial_X \psi \right) \right \} \right ] .
\end{equation}
$\delta^2 f_{\rho}^{\rm L}$ vanishes only if $I'=0$ and $W= \Omega - \Omega_{\rm F}=0$.
When $I'$ and $W$ are finite, $\delta^2 f_{\rho}^{\rm L}$ is finite and
the force-free condition is broken.
% Since $\delta^2 f_0^{\rm L} = 0$ and Eq. (\ref{eneconeqconaco}), the energy is conserved,
% while the angular momentum is not conserved in the corotating natural coordinates 
% because of $\delta^2 f_{\rho}^{\rm L} \neq 0$ and Eq. (\ref{anmconeqconaco}).
% The non-conservation of the angular momentum in the corotating natural coordinates
% yields the violation of the 
% conservation of the energy in the coordinates $(t, r, \varphi, z)$
% as shown in subsection \ref{subsec3}.
% In fact, as shown by the numerical simulations in Section \ref{secmet},
% the conservation of energy calculated only with $\psi$ is broken.
Then, the conservation law of the energy and angular momentum is not hold
as far as we consider the Alfv\'{e}n wave only.
This breakdown of the conservation law is apparent and is recovered 
by the additional wave as shown in the next subsection.

\subsection{Recovery of the force-free condition
\label{subsec2}}

To recover the force-free condition, we should take into account
of the additional first and second-order perturbations $\delta^2 \phi_1$, 
$\delta \phi_2$, $\delta^2 \phi_2$ as
%%% Last, we have to guarantee the second order of Lorentz force of the perturbation
%%% with respect to the Alfv\'{e}n wave. To discuss the second order of the perturbation,
%%% we have to consider the expansion of the perturbation up to second order,
\begin{eqnarray}
\phi_1 &= \bar{\phi}_1 + \delta \phi_1 + \delta^2 \phi_1 = -z + \psi + \delta^2 \phi_1, \\
\phi_2 &= \bar{\phi}_2 + \delta \phi_2 + \delta^2 \phi_2 
= \rho + \delta \phi_2 + \delta^2 \phi_2.
\end{eqnarray}
We should notice that the force-free condition is broken only in one
equation with respect to $\delta^2 J_\rho$. Then, the breakdown of 
the force-free condition should be recovered by addition of 
one freedom variable. The above intuitive reason of the Lorentz force acting on
the Alfv\'{e}n wave suggests the variable $\delta^2 \phi_2$ is the appropriate
additional perturbation as the additional freedom.
%% In fact, when we introduce the variable $\chi=\delta^2 \phi_1$, we can 
To recover the force-free condition 
%% as shown in the next section.
%% \section{Energy conservation of Alfv\'{e}n wave with force-free condition recovered by $\chi$
%% \label{sec4}}
we consider the following perturbation setting of the force-free Alfv\'{e}n wave
along the magnetic field line around the spinning black string:
%% for the force-free electromagnetic field:
\begin{eqnarray}
\phi_1 &=& \bar{\phi}_1 + \delta \phi_1 = -z + \psi(T, X), \\
\phi_2 &=& \bar{\phi}_2 + \delta^2 \phi_2 = \rho + \chi(T, X) ,
\end{eqnarray}
where $\chi \equiv \delta^2 \phi_2$.

The time evolution equation of $\chi$ is given as
\begin{equation}
\Box \chi = \nabla_\mu \nabla^\mu \chi = \frac{1}{\sqrt{-g}} \partial_\lambda
(\sqrt{-g} g^{\mu\beta} \partial_\beta \chi)
= \frac{1}{\sqrt{-g}} \partial_\lambda (\sqrt{-g} W^{\lambda\alpha\mu\rho}
\partial_\alpha \psi \partial_\mu \psi) \equiv s,
\label{eqchi4dlb_tx}
\end{equation}
where $s$ is the recognized as the source term in the wave equation
with respect to $\chi$ (see the derivation in Appendix \ref{appendb}).
Eq. (\ref{eqchi4dlb_tx}) reads
\begin{equation}
\Box \chi = \frac{1}{\sqrt{-g}} \left [
\sqrt{-g} g^{TT} \partial_T^2 \chi + 2 f \partial_x ( \sqrt{-g} g^{TX} \partial_T \chi )
- \partial_x(\sqrt{-g}) \partial_T \chi + f \partial_x (\sqrt{-g} g^{XX} f \partial_x \chi)
\right ] = s.
\label{chieq}
\end{equation}
When we consider $\chi$, we confirm that the force-free condition holds
as shown in Appendix \ref{appendb};
eventually, introducing $\chi$, we recover and confirm the force-free condition.
The linear homogeneous equation (Eq. (\ref{eqchi4dlb_tx}))
shows the wave described by $\chi$
propagates isotropically in space, and then, it is recognized by the 
fast wave \citep{bellman06},
while the complex formalism of homogeneous linear equation of $\psi$ 
(Eq. (\ref{eq4i2})) suggests
the strong anisotropy of the wave described by $\psi$, 
which is characteristic property of the Alfv\'{e}n wave.

%%% \subsection{Energy and angular momentum conservation law 
%%% in the corotating natural coordinates}

% In the corotating natural coordinates, $f_T^{\rm L}$ vanishes, then
% the energy conservation law stands even without introducing $\chi$.
% However, including $\chi$, we have to recalculate the energy density
% and energy flux.
The equilibrium and first order of these values are the same
as the calculation without introducing $\chi$.
%% We summarize the up to second order of the energy density and energy flux
When $\xi^\mu$ is the Killing vector, we have the conservation law,
\begin{equation}
\frac{\partial}{\partial x^0} \xi_\nu T^{0 \nu}
+ \frac{1}{\sqrt{-g}} \frac{\partial}{\partial x^i} (\sqrt{-g} \xi_\nu T^{i \nu})
= \frac{1}{\sqrt{-g}} \partial_\mu (\sqrt{-g} T^{\mu\nu}) 
= \nabla_\mu (\xi_\nu T^{\mu\nu}) = - \xi^\nu f^{\rm L}_\nu .
\end{equation}
When $\xi^\nu f^{\rm L}_\nu$ vanishes,
$\xi_\nu T^{0 \nu}$ represents the density of conservation value
and $\xi_\nu T^{i \nu}$ represents the density of the conservation quantity flux.
% If $\xi^\nu f^{\rm L}_\nu$ is finite, we have no conservation with respect
% to the Killing vector $\xi^\mu$.
% Then, to consider the conservation laws of the electromagnetic energy
% and angular momentum, we have to guarantee the force-free condition
% of the electromagnetic field.

% \paragraph{Energy transport due to Alfv\'{e}n wave}

For the time-like Killing vector $\xi^\mu_{(T)} = (1, 0, 0, 0)$ 
and the axial Killing vector $\xi^\mu_{(\rho)} = (0, 0, 1, 0)$ 
in the corotating natural frame,
we have the conservation laws of the energy and angular momentum 
\begin{eqnarray}
\frac{\partial}{\partial x^0} S^0
& + & \frac{1}{\sqrt{-g}} \frac{\partial}{\partial x^i} (\sqrt{-g} S^i) = f_0^{\rm L} ,\\
\frac{\partial}{\partial x^0} M^0
& + & \frac{1}{\sqrt{-g}} \frac{\partial}{\partial x^i} (\sqrt{-g} M^i) = - f_\rho^{\rm L} ,
\end{eqnarray}
where $S^\mu = - \xi^{(T)}_\nu T^{\mu\nu}$ is 4-energy flux
and $M^\mu = \xi^{(\rho)}_\nu T^{\mu\nu}$ is 4-angular momentum flux.
%%% if $f_\mu^{\rm L}$ vanishes in the force-free case.
Furthermore, when we assume the axisymmetry and $z$-direction translation symmetry,
we have
\begin{eqnarray}
\frac{\partial}{\partial x^0} S^0
&+& \frac{1}{\sqrt{-g}} \frac{\partial}{\partial X} (\sqrt{-g} S^X) = f_0^{\rm L}, 
\label{eneconeqconaco} \\
\frac{\partial}{\partial x^0} M^0
&+& \frac{1}{\sqrt{-g}} \frac{\partial}{\partial X} (\sqrt{-g} M^X) = -f_\rho^{\rm L},
\label{anmconeqconaco} 
\end{eqnarray}

Since the energy density and flux $S^0$ and $S^X$ are calculated as
$S^0 = -\frac{1}{2} F^{0i} F_{0i} + \frac{1}{4} F^{ij} F_{ij}$,
$S^X = F^{Xi} F_{i0}$, respectively,
we have the energy density and flux of the 
% equilibrium, first ($O(\psi)$) and 
second-order perturbation
% ($O(\psi \times \psi)$) 
of the linear Alfv\'{e}n wave and the induced fast wave
% given by the solution $\psi$ of Eq. (\ref{newtontoyeq}),
%%% \begin{eqnarray}
%%% &\displaystyle \bar{S}^0 = \frac{I'^2 - \Gamma}{2 X^2 {\alpha}^2} , &
%%% \bar{M}^0 = - \frac{\Gamma - I'^2}{\alpha^2 \Gamma} (\Omega - \Omega_{\rm F}) \\
%%% & \delta S^0 = 0, &
%%% \delta M^0 = 0, 
%%% \label{delsdelmvan} \\
%%% &\displaystyle \delta^2 S^0 =
%%% \frac{-\gamma_{\rm F}^2}{2 \Gamma} [\lambda (\partial_T \psi)^2 +(\partial_x \psi)^2],&
%%% \verb!   !
%%% \delta^2 M^0 
%%% = - \partial_X \psi \left [ (\Omega - \Omega_{\rm F}) \partial_X \psi
%%% + \frac{I'}{X \Gamma} \partial_T \psi \right ],
%%% \end{eqnarray}
%%% where the detailed derivation is shown in Appendix \ref{appenda}.
(see detailed derivation in Appendix \ref{appendb}) as follows:
\begin{eqnarray}
\delta^2 S^0 &=& 
% \frac{-\gamma_{\rm F}}{2 \Gamma} [\lambda (\partial_T \psi)^2 +(\partial_x \psi)^2]
% + \frac{I' \gamma_{\rm F}}{-\Gamma} \partial_x \chi. \\
\frac{1}{2} \left [
\frac{I'^2 - \Gamma}{- \Gamma \alpha^2 X^2} (\partial_T \psi)^2
+ \frac{-\Gamma}{X^2} (\partial_X \psi)^2 \right ]
+ \frac{I'}{X} \partial_X \chi , \\
\delta^2 S^X &=&
% - \frac{\gamma_{\rm F}}{X} \partial_T \psi \partial_x \psi 
\frac{\Gamma}{X^2} \partial_T \psi \partial_X \psi
- \frac{I'}{X} \partial_T \chi.
\label{d2sx}
\end{eqnarray}

Since the angular momentum density and flux are 
$M^0=F^{0 \nu} F_{\rho\nu}$ and $M^X = F^{X\nu} F_{\rho\nu}$,
we have 
% the energy and angular momentum fluxes 
% of the equilibrium, first and
% second-order perturbations of the linear Alfv\'{e}n wave as follows
% \begin{eqnarray}
% & \displaystyle \bar{S}^X =  0, & \bar{M}^X = \frac{I'}{X}  \\
% &\delta S^X = 0, & \delta M^X = 0, \\
% & \delta^2 S^X = - \frac{\gamma_{\rm F}}{X} \partial_T \psi \partial_x \psi ,& 
% \verb!   !
% \delta^2 M^X 
% = \partial_T \psi \left [ (\Omega - \Omega_{\rm F} ) \partial_X \psi
% + \frac{I'}{X \Gamma} \partial_T \psi \right ].
% \end{eqnarray}
% Eq. (\ref{d2mmuwg}) yields 
the angular momentum density and flux of the second perturbation as follows
\begin{eqnarray}
\delta^2 M^0 &=& 
- \partial_X \psi \left [ (\Omega - \Omega_{\rm F}) \partial_X \psi
+ \frac{I'}{X \Gamma} \partial_T \psi \right ]
- \frac{\gamma_{\rm N}}{X^2} \partial_T \chi, \\
\delta^2 M^X &=&
\partial_T \psi \left [ (\Omega - \Omega_{\rm F}) \partial_X \psi
+ \frac{I'}{X \Gamma} \partial_T \psi \right ]
+ \frac{I'X}{\Gamma} (\Omega - \Omega_{\rm F}) \partial_T \chi
+ \alpha^2 \partial_X \chi,
\end{eqnarray}
where the detail calculation is shown in Appendix \ref{appendb}.

%%% We have the energy density and energy flux density of the equilibrium, first and
%%% second-order perturbations of the linear Alfv\'{e}n wave distinctively
%%% (see Appendix \ref{appendb}),
%%% \begin{eqnarray}
%%% \bar{M}^0 &=& g^{0 \rho} = - \frac{\Gamma - I'^2}{\alpha^2 \Gamma} (\Omega - \Omega_{\rm F}), \verb!   !
%%% \bar{M}^X = g^{X\rho} = \frac{I'}{X}, \\
%%% \delta M^0 &=&  0, \verb!   !
%%% \delta M^X = 0, \\
%%% \delta^2 M^0 &=& 
%%% - W^{T\lambda X \rho} \partial_\lambda \psi \partial_X \psi 
%%% + g^{T \beta} \partial_\beta \chi ,\\
%%% \delta^2 M^X &=& 
%%% W^{T\lambda X \rho} \partial_\lambda \psi \partial_T \psi 
%%% + g^{X \beta} \partial_\beta \chi .
%%% \end{eqnarray}

% These expansions show the second-order perturbation should be included
% to investigate the energy transport of the linear Alfv\'{e}n wave.
% Then, the force-free condition should be satisfied up to second order 
% of the perturbation. If the force-free condition of the second-order 
% perturbation is broken, the energy conservation of the Alfv\'{e}n wave
% does not hold.
% The conservation of the angular momentum of the Alfv\'{e}n wave is the same with
% energy conservation of the Alfv\'{e}n wave.
% To recover the conservation of the energy and the angular momentum, we checked
% the force-free condition up to the second order of the perturbation
% of the Alfv\'{e}n wave.

\subsection{Conservation laws of energy in the BTZ spacetime
\label{subsec3}}
Now, we derive the energy density and energy flux for the energy conservation law
in the BTZ coordinates.
In this subsection, we have to use both the original coordinates
of the BTZ spacetime $(t, r, \phi, z)$
and the corotating natural coordinates $(T, X, \varphi, z)$.
To distinguish the two coordinates,
we denote the corotating natural coordinates by $x^{\underline{\mu}}=(T, X, \rho, z)$
with the underline, while
the original coordinates of the BTZ spacetime just by $x^\mu = (t, r, \varphi, z)$.
First, we formulate the energy conservation in the BTZ coordinates.
The Killing vector for the energy conservation law in the BTZ coordinates is
$\xi_{(t)}^{{\mu}}=(1, 0, 0, 0)$.
Then, the 4-energy flux density is 
\begin{equation}
{S}^{{\mu}} = - \xi_{(t)}^{{\nu}} 
T^{{\mu}}_{{\nu}} = - T^{{\mu}}_t
= - T^{{\mu}}_t.
\end{equation}
% where the underlines of the quantities indicates the variable in the
% BTZ coordinate system.
The energy conservation law is expressed by
\begin{equation}
\nabla_{{\mu}} {S}^{{\mu}}
= \frac{1}{\sqrt{-g}} \partial_{\mu} 
(\sqrt{-g} {S}^{{\mu}})
= \xi_{(t)}^{{\nu}} {f}_{{\nu}}^{\rm L} 
= {f}_t^{\rm L}
= \frac{\partial x^{\underline{\mu}}}{\partial t} f_{\underline{\mu}}^{\rm L} = 0,
\end{equation}
after we introduced $\chi$.
The energy density in the BTZ coordinates is related with the components
of density and flux in the corotating natural coordinates as
\begin{equation}
{S}^t  =  
\underline{S}^T + \Omega_{\rm F} \underline{M}^T - I_0 (\underline{S}^X 
+ \Omega_{\rm F} \underline{M}^X),
\end{equation}
where detailed derivation is shown in Appendix \ref{appendc}.
Then, we have the energy density and energy flux density of the equilibrium, first and
second-order perturbations with respect to the linear Alfv\'{e}n wave
and the second-order fast wave
(see derivation in Appendix \ref{appendc}),
\begin{eqnarray}
{\bar{S}}^t &=& 
\frac{1}{2} g^{\rho\rho} + \Omega_{\rm F} (g^{T\rho} - I_0 g^{X \rho}), \\
\delta {S}^t &=& 0 , \\
\delta^2 {S}^t &=&
\delta^{1+1} {S}^t +  \delta^{2+0} {S}^t  ,
\label{d2st}
\end{eqnarray}
where
\begin{eqnarray}
\delta^{1+1} {S}^t 
%% & = & -\frac{1}{2} W^{TT\rho\rho} (\partial_T \psi)^2 
%% +\frac{1}{2} W^{XX\rho\rho} (\partial_X \psi)^2 
%% -  \Omega_{\rm F} W^{T\lambda X \rho} \partial_\lambda \psi \partial_X \psi
%% + I_0 (W^{XX\rho\rho} \partial_X \psi \partial_T \psi + \Omega_{\rm F} 
%% W^{X\nu\rho\rho} \partial_\nu \psi \partial_T \psi) \nonumber \\
& = & \frac{1}{2 \Gamma^2} \left [ \frac{1}{X^2 {\alpha}^2} \left \{ 
\Gamma^2 + I'^2({\alpha}^2 - X^2 (\Omega^2 - \Omega_{\rm F}^2)) \right \} (\partial_T \psi)^2
+ \frac{\Gamma^2}{X^2} \left \{ {\alpha}^2 - X^2 (\Omega^2 - \Omega_{\rm F}^2) \right \} 
(\partial_X \psi)^2 \right . \nonumber \\
& &  \left . - 2 \frac{\Gamma}{{\alpha}^2 X} I'
\{ \Gamma (\Omega - \Omega_{\rm F}) + \Omega_{\rm F}
({\alpha}^2 + X^2 (\Omega - \Omega_{\rm F})^2 ) \} \partial_T \psi \partial_X \psi  \right ]
 , 
\label{d11st} \\
\delta^{2+0} {S}^t & = & g^{X \rho} \partial_X \chi + \Omega_{\rm F} 
g^{T \beta} \partial_\beta \chi + I_0 (g^{X \rho} \partial_T \chi 
- \Omega_{\rm F} g^{X \beta} \partial_\beta \chi) \nonumber \\
& = & \frac{I'}{X} \partial_X \chi - \frac{1}{{\alpha}^2 \Gamma}
(\Omega_{\rm F} \Gamma - I'^2 W) \partial_T \chi.
\label{d20st}
\end{eqnarray}
In the numerical calculations, we use the above formulation
(Eqs. (\ref{d2st})-(\ref{d20st})) with the tortoise 
coordinates.
% When we use the tortoise coordinate $x$, we have the following expressions,
% \begin{eqnarray}
% \delta^{1+1} {S}^t & = & 
% \frac{1}{2 \Gamma^2} \left [ \frac{1}{X^2 {\alpha}^2} \left \{ 
% \Gamma^2 + I'^2 ({\alpha}^2 - X^2 (\Omega^2 - \Omega_{\rm F}^2)) \right \} (\partial_T \psi)^2
% + \gamma_{\rm F}^2 \left \{ {\alpha}^2 - X^2 (\Omega^2 - \Omega_{\rm F}^2) \right \} 
% (\partial_x \psi)^2 \right . \nonumber \\
% & &  \left . + 2 \frac{I' \gamma_{\rm F}}{{\alpha}^2} 
% \{ \Gamma (\Omega - \Omega_{\rm F}) + \Omega_{\rm F}
% ({\alpha}^2 + X^2 (\Omega - \Omega_{\rm F})^2 ) \} \partial_T \psi \partial_X \psi  \right ]
%  , \\
% \delta^{2+0} {S}^t & = & 
% \frac{\gamma_{\rm F} I'}{- \Gamma} \partial_x \chi - \frac{1}{{\alpha}^2 \Gamma}
% (\Omega_{\rm F} \Gamma - I'^2 W) \partial_T \chi.
% \end{eqnarray}

The energy flux density in the BTZ coordinates is related with the flux components
in the corotating natural coordinates by
\begin{equation}
{S}^r  = 
 - T^X_T + \Omega_{\rm F} T^X_\rho = \underline{S}^X + \Omega_{\rm F} \underline{M}^X,
\end{equation}
where detail derivation is shown in Appendix \ref{appendc}.
Then, we have the energy flux density of the equilibrium, first and
second-order perturbations of the linear Alfv\'{e}n wave
(see Appendix \ref{appendc}),
\begin{eqnarray}
{\bar{S}}^r &=& 
\Omega_{\rm F} \frac{I'}{X} ,\\
\delta {S}^r &=&  0, \\
\delta^2 {S}^r &=& \delta^{1+1} {S}^r +  \delta^{2+0} {S}^r  ,
\label{d2sr}
\end{eqnarray}
where
\begin{eqnarray}
\delta^{1+1} {S}^r & = &
- \frac{1}{X^2} \partial_T \psi \left [ 
\left (X^2 - M + \frac{Ma}{2} \Omega_{\rm F} \right ) \partial_X \psi
- \frac{I' X \Omega_{\rm F}}{\Gamma} \partial_T \psi  \right ] , \\
\delta^{2+0} {S}^r & = &
 \frac{I'}{X \Gamma} ( - \Gamma + X^2 W \Omega_{\rm F} ) \partial_T \chi
+ \Omega_{\rm F} \underline{\alpha}^2 \partial_X \chi .
\label{d20sr}
\end{eqnarray}
% When we use the tortoise coordinate $x$, we have distinct expressions,
% \begin{eqnarray}
% \delta^{1+1} {S}^r & = & - \frac{\gamma_{\rm F}}{X \Gamma} 
% \partial_T \psi \left [ 
% \left (X^2 - M + \frac{Ma}{2} \Omega_{\rm F} \right ) \partial_x \psi
% + \frac{I' \Omega_{\rm F}}{\gamma_{\rm F}} \partial_T \psi  \right ] , \\
% \delta^{2+0} {S}^r & = & 
% \frac{I'}{X \Gamma} [
% {\alpha}^2 - X^2 (\Omega - \Omega_{\rm F})(\Omega - 2 \Omega_{\rm F}) ]  \partial_T \chi
% - {\alpha}^2 \frac{\Omega_{\rm F} X}{\Gamma} \gamma_{\rm F} \partial_x \chi.
% \end{eqnarray}
In the numerical calculations, we use the formulation 
(Eqs. (\ref{d2sr})--(\ref{d20sr}))
with the tortoise coordinates.

%% \section{Energy conservation in the BTZ spacetime}
%% \label{sec5}}

\section{Numerical method
\label{secmet}}

To perform 1D numerical simulations of the force-free field with Eq. (\ref{nodaeq19d}), 
we use the multi-dimensional two-step Lax-Wendroff scheme of 
$\VEC{u} = (\psi, v, w, \chi, u, G)$ whose details
are shown in Appendix \ref{appendnum}.
Here, $v, w, \chi, u$, and $G$ 
are the new variables introduced for the numerical calculation.
This scheme sometimes causes numerically artificial structure called
the Gibbs phenomena. To avoid such numerical phenomena, we employ
a smooth profile of initial variables as follows.
%
% \paragraph{Initial condition}
The initial condition of a outwardly/inwardly propagating single pulse of field is given as:
\begin{eqnarray}
\psi(x) &\equiv& \left \{ 
\begin{array}{cc} \displaystyle \left [ 1 - \left | \frac{x-x_0}{w/2} \right |^{2m} \right ]^n
& \displaystyle \left (x_0 - \frac{w}{2} \leq x \leq x_0 + \frac{w}{2} \right ) \\
% 1 & (x < x_0) \\
0 & ({\rm other}) \end{array} \right . \nonumber \\
v(x) &=& \pm \psi(x) \\
w(x) &=& 0 \nonumber \\
\chi(x) &=& 0 \nonumber \\
u(x) &=& 0 \nonumber \\
G(x) &=& 0 \nonumber
\end{eqnarray}
where the plus and minus signs in the equation with respect to 
$v(x)$ are taken for the outwardly and
inwardly propagating waves, respectively,
$x_0$ and $w$ give the center and the width of the pulse of the Alfv\'{e}n wave and $m$ and $n$ 
are constants of the pulse shape. In this paper, we set
$n=32$ and $m=\log(1-2^{-1/32})/\log(1-2^{-1/4})$.
Hereafter, through the sections of numerical calculation, we set $M$ unity.

%% With respect to a inwardly propagating single pulse of field, the initial
%% condition is given by
%% \begin{eqnarray}
%% \psi(x) &\equiv& \left \{ 
%% \begin{array}{cc} \displaystyle \left [ 1 - \left | \frac{x-x_0}{w/2} \right |^{2m} \right ]^n
%% & \displaystyle \left (x_0 - \frac{w}{2} \leq x \leq x_0 + \frac{w}{2} \right ) \\
%% 0 & ({\rm other}) 
%% % 1 & (x > x_0) 
%% \end{array} \right . \nonumber \\
%% v(x) &=& -\psi(x) \\
%% w(x) &=& 0 \nonumber \\
%% \chi(x) &=& 0 \nonumber \\
%% u(x) &=& 0 \nonumber \\
%% G(x) &=& 0 \nonumber
%% \end{eqnarray}
As shown in the next section, we have very smooth numerical results
without numerical artificial structure.
In this paper, we set the width and the position of the outwardly 
and inwardly propagating pulses by $w=0.5$, $x_0=-1.6$ (outward pulse)
and $w=3$, $x_0=2$ (inward pulse), respectively.

At the inner and outer boundaries $x=x_{\rm min}$ and $x=x_{\rm max}$,
the free boundary conditions, $\VEC{u}_0 - \VEC{u}_1 = \VEC{0}$,
$\VEC{u}_{I} - \VEC{u}_{I-1} = \VEC{0}$
% $\displaystyle \frac{\partial \VEC{u}}{\partial x} = \VEC{0}$ 
are used to mimic the radial boundary condition,
where $\VEC{u}_{0}$ and $\VEC{u}_{I}$ are the variables
at the boundary and $\VEC{u}_{1}$ and $\VEC{u}_{I-1}$ are
the values at the neighborhood.

\section{Numerical results 
\label{secres}}

\begin{figure} % [H]
\begin{center}
\includegraphics[width=17cm]{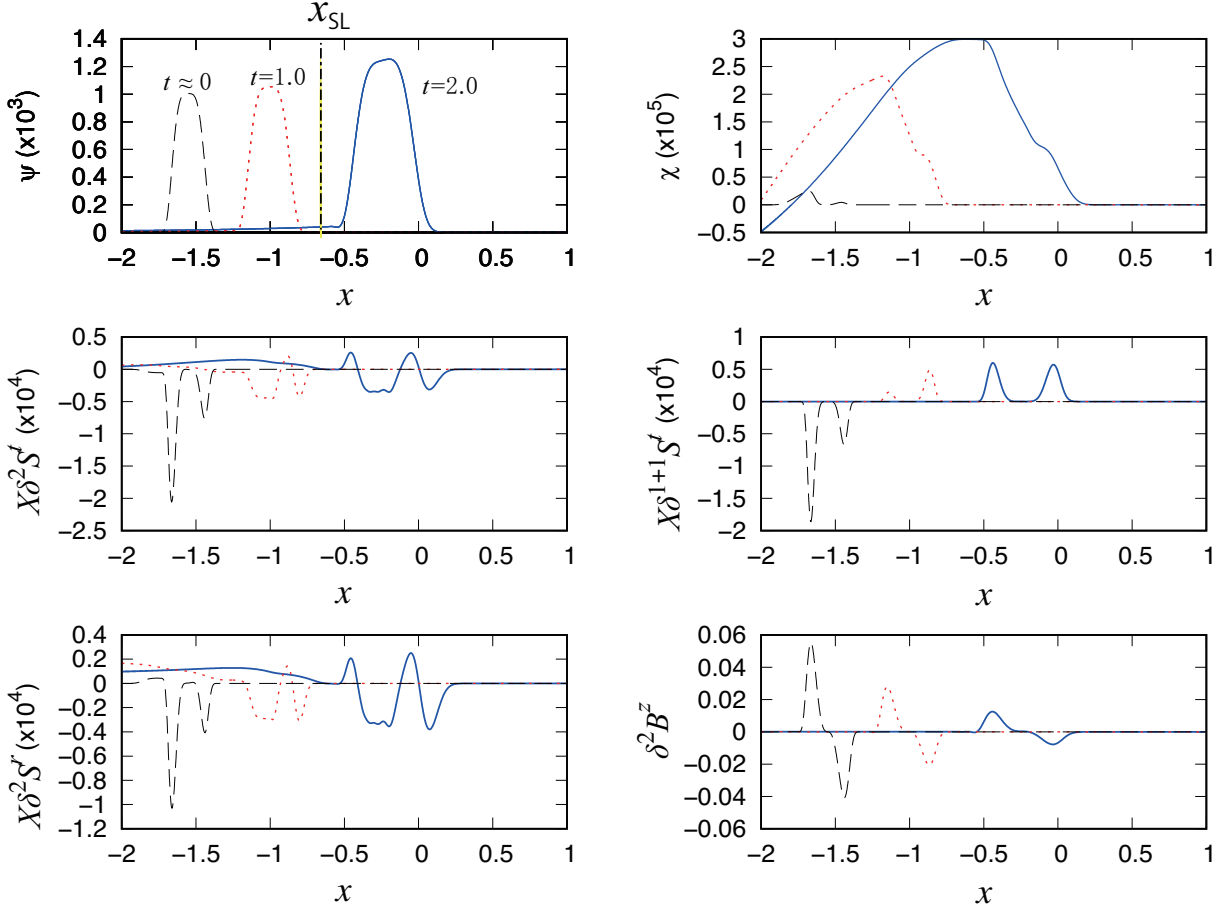}% Here is how to import EPS art
\end{center}
\caption{Time evolution of the variables of a force-free electromagnetic
field for an outwardly propagating pulse of the Alfv\'{e}n wave from a region
in the ergosphere along a magnetic field line of $\Omega_{\rm F} =0.5$
(the case of black string energy extraction by the Blandford--Znajek mechanism)
%%  because $0 < \Omega_{\rm F} < \Omega_{\rm H}$)
around a spinning black string with the spin parameter $a=0.9$ at
$t=$ 0.062 (black dashed line), 1.0 (red dotted line), 2.0 (blue, thick solid line).
The dashed-dotted line shows the location of the static limit point.
$\sqrt{-g} \delta^2 {S}^t = X \delta^2 {S}^t$ is the total energy
density of the Alfv\'{e}n wave and the
induced wave, which oscillates azimuthally. 
$\sqrt{-g} \delta^{1+1} {S}^t = X \delta^{1+1} {S}^t$
is the total energy density of the Alfv\'{e}n wave only, which is calculated
from $\delta^{1+1} {S}^t$ of $\psi$ without $\chi$. 
$\delta B^z$ is the $z$-component of the
magnetic field observed in the natural coordinates,
$\displaystyle \delta B^z = \frac{1}{\sqrt{\gamma_{\rm N}}} \partial_X \psi
= \frac{X e^{-2 x}}{\sqrt{\gamma_{\rm N}}} \partial_x \psi$.
\label{fig_ow_times}}
\end{figure}

\begin{figure} % [H]
\begin{center}
\includegraphics[width=17cm]{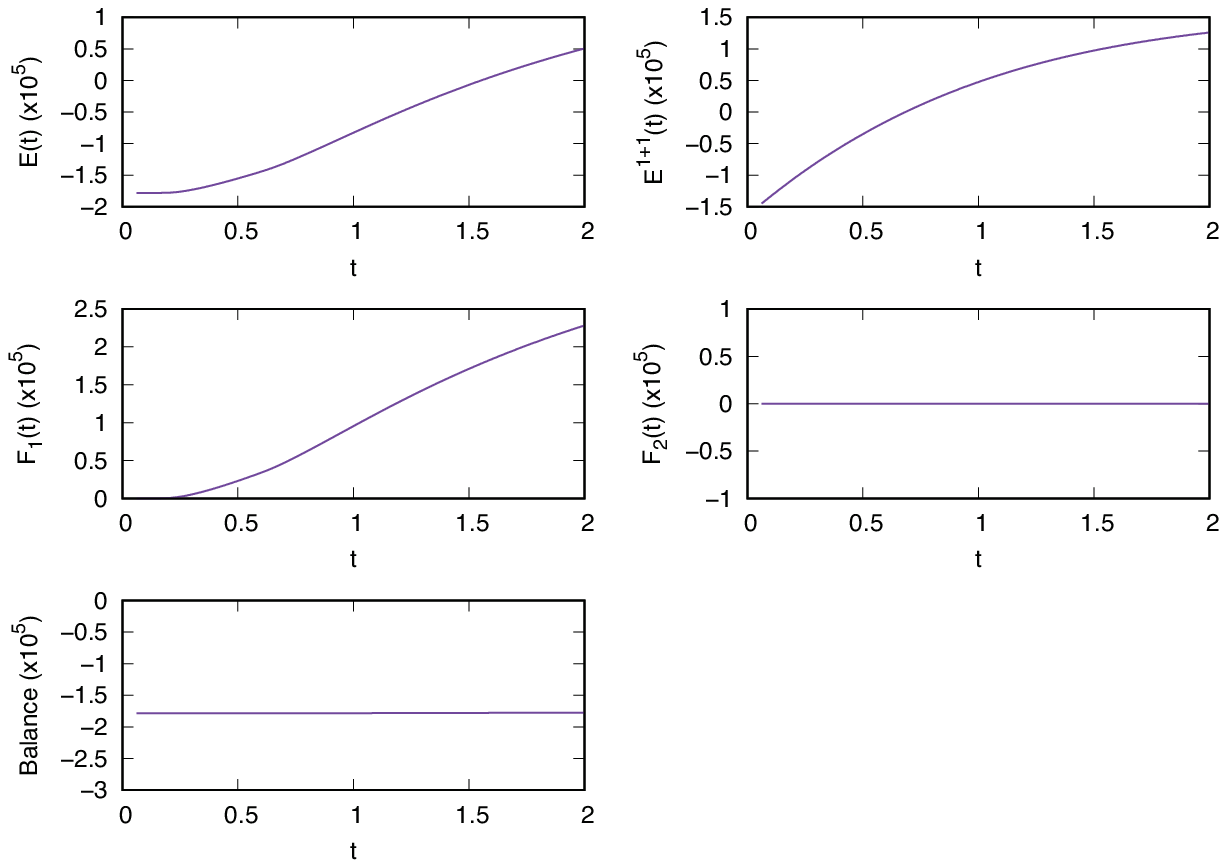}% Here is how to import EPS art
\end{center}
\caption{Time evolution of the energy balance of the Alfv\'{e}n 
and induced fast waves.
$E(t)$: total energy. 
$E^{1+1}(t)$: energy of the Alfv\'{e}n wave only. 
$F_1(t)$: energy flux at the left edge.
$F_2(t)$: energy flux at the right edge.
The balance: $E(t) - F_1(t) + F_2(t)$.
The energy balance constancy shows the energy conservation of the
Alfv\'{e}n and induced fast waves.
\label{fig_ow_his5t}}
\end{figure}

We performed the 1D numerical simulations of Alfv\'{e}n waves
along a stationary magnetic field line with $\Omega_{\rm F}=0.5$
($0 < \Omega_{\rm F} < \Omega_{\rm H}$) 
and $\Omega_{\rm F}=0.7$ ($\Omega_{\rm F} > \Omega_{\rm H}$) at $z=0$ 
around a black string with spin parameter $a=0.9$ using Eqs. (\ref{nodaeq19d})
and (\ref{chieq}).
%% In the numerical simulation, we set $\mu^2=0$.
In the case, we have the horizon radius $r_{\rm H}=\sqrt{(1+\sqrt{1-a^2})/2}=0.8473$,
the radius of the light surface $r_{\rm LS}=0.8563$, $x$ of the static limit surface
($r_{\rm SL} = 1$), $x_{\rm SL}
=\frac{1}{2} \log [\Omega_{\rm F} ( a - \Omega_{\rm F})/(1-\Omega_{\rm F}^2)]=-0.661$
in the case of $\Omega_{\rm F}=0.5$ and $r_{\rm LS}=0.852$,
$x_{\rm SL}=-0.6464$ in the case of $\Omega_{\rm F}=0.7$. 
%%% Here, we set $M$ unity.

The initial condition of $\psi$ is given by $w=0.5$ and $x_0 = -1.5$ at $T=0$ in the case
of outwardly propagating pulse both in the background magnetic field with
$\Omega_{\rm F} = 0.5$ and $\Omega_{\rm F} = 0.7$.
In the case of the inwardly propagating pulse, we set $w=3$, $x_0=4$, 
and $\Omega_{\rm F} = 0.5$ for the initial condition of $\psi$ at $T=0$.
The initial condition of $\chi$ is given by $\chi = 0$ in all cases at $T=0$.

Figure \ref{fig_ow_times} shows the Alfv\'{e}n wave propagates outwardly and the
fast induced wave is caused by the Alfv\'{e}n wave
and propagates toward the both sides in the case of $\Omega_{\rm F} = 0.5$,
$a=0.9$, $w=0.5$, and $x_0=-1.5$. 
The lines in each panel show the time shot at $t=0.062$ (black dashed line),
$t=1.0$ (red dotted line), and $t=2.0$ (blue, thick solid line).
The energy density of the Alfv\'{e}n wave
$X \delta^{1+1} S^t$
is initially negative but at the outside of the ergosphere ($X>1$), 
the energy density becomes positive, where
the total energy density of the Alfv\'{e}n wave
$\sqrt{-g} \delta^{1+1} {S}^t = X \delta^{1+1} {S}^t$
is calculated from only $\delta^{1+1} {S}^t$ of $\psi$ without $\chi$
(see detail in Appendix \ref{appendetbtz}). 
The total energy density of the Alfv\'{e}n wave
and the induced wave, $X \delta^{1+1} S^t$,
remains dominantly negative. The integrated energy over
the calculation region ($E(t)$ in Eq. (\ref{eqofet})) is shown in Fig. 5. 
The total energy balance of the Alfv\'{e}n
wave and the induced wave remains constant.
This shows that the total energy of the Alfv\'{e}n wave and the induced wave is conserved.
% In the last stage, the balance is broken
% because the data is lost at the left edge region 
% due to the frame transformation from the natural coordinates to the BTZ coordinates.
On the other hand, the integrated energy of the Alfv\'{e}n wave 
over the calculation region increases
monotonically from negative value to positive as shown in Fig. \ref{fig_ow_his5t}. 
This shows the Alfv\'{e}n wave obtains the energy from the background magnetic field.
This phenomena is explained by the relativistic mechanism, where
the Alfv\'{e}n wave has the angular momentum.
%% , in section \ref{sec3}.
% This result is reasonable. Because it is explained intuitively as follows.
In the framework of relativity,
the Alfv\'{e}n wave propagating along the magnetic field line has its angular momentum. 
Then, if the magnetic field line is curved ($I' \ne 0$),
the torque onto the
Alfv\'{e}n wave is required to trace the magnetic field line.
This means the force-free condition is broken and the azimuthal component
of the Lorenz force becomes finite.
Even if the magnetic field line is straight, the propagation direction
of the Alfv\'{e}n wave changes when the magnetic field line rotates 
($W = \Omega - \Omega_{\rm F} \ne 0$). That is, the angular momentum
of the Alfv\'{e}n wave changes. Then, the Alfv\'{e}n wave should receive external 
torque to trace the rotating magnetic field line.
Even in the case of the straight magnetic field line, 
the force-free condition is also broken when the magnetic field rotates.
Now, we confirmed that $\chi$ recovers the energy conservation law in the BTZ coordinates
as expected.

\begin{figure} % [H]
\begin{center}
\includegraphics[width=17cm]{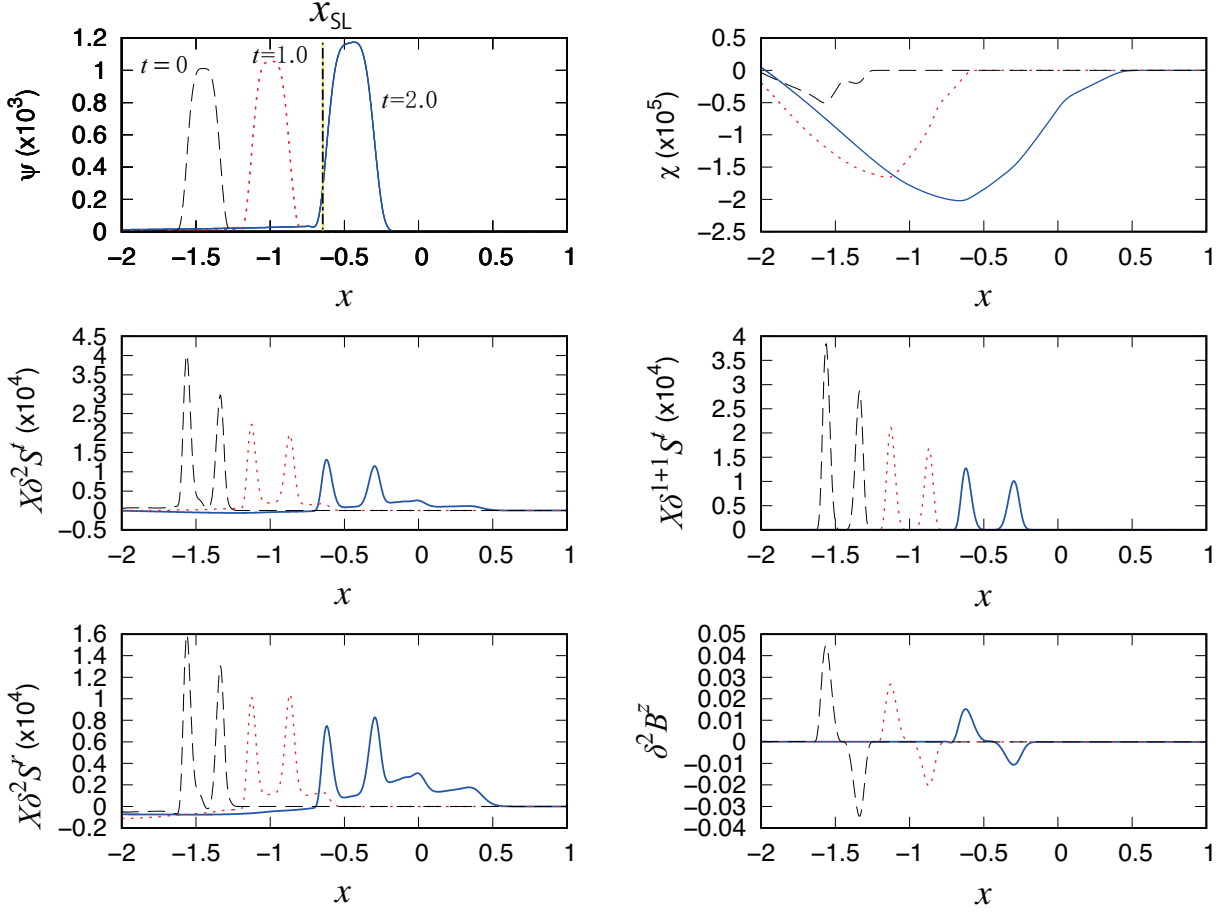}% Here is how to import EPS art
\end{center}
\caption{Similar to Fig. \ref{fig_ow_times}, but for the calculation
of the case with the background magnetic field of $\Omega_{\rm F} = 0.7$
($\Omega_{\rm F} > \Omega_{\rm H}$)
at $t=$ 0 (black dashed line), 1.0 (red dotted line), 2.0 (blue, thick solid line).
.
\label{fig_ow_plt6t_bzof}}
\end{figure}

\begin{figure} % [H]
\begin{center}
\includegraphics[width=17cm]{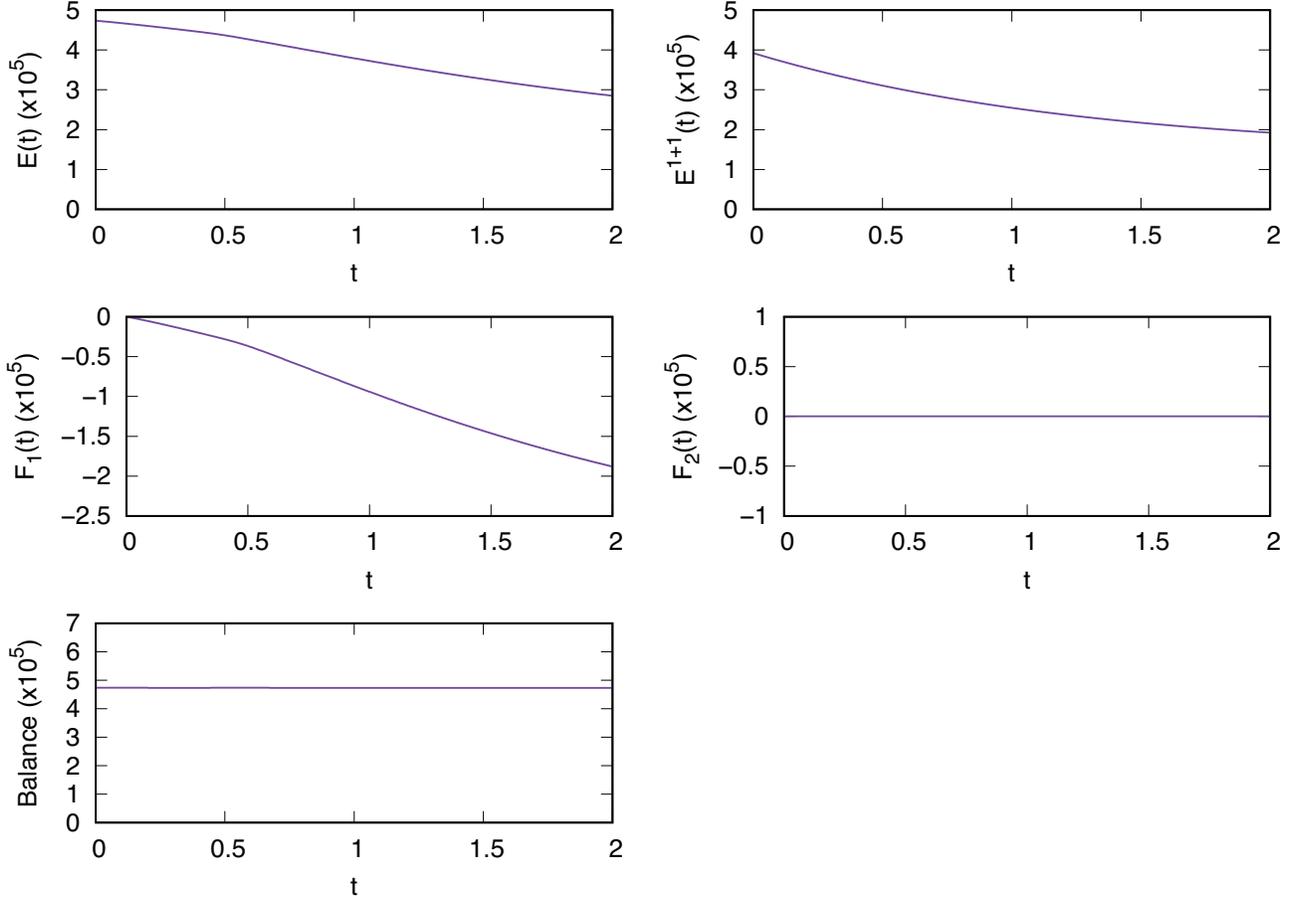}% Here is how to import EPS art
\end{center}
\caption{Similar to Fig. \ref{fig_ow_his5t}, but for the calculation of the case
with the background magnetic field $\Omega_{\rm F} = 0.7$
($\Omega_{\rm F} > \Omega_{\rm H}$).
\label{fig_ow_his5t_bzof}}
\end{figure}

As shown in Fig. \ref{fig_ow_plt6t_bzof}, 
we also performed a calculation of the Alfv\'{e}n wave along the magnetic field
line of $\Omega_{\rm F} =0.7$, where the rotational energy of the black string
is not extracted by the Blandford--Znajek mechanism.
The Alfv\'{e}n wave propagates outwardly and the fast wave is induced.
The induced wave propagates toward both sides as the case of $\Omega_{\rm F}= 0.5$
(the Blandford--Znajek mechanism extracts the black string rotational energy). 
However, the amplitude of the induced
wave is so small compared to the case of $\Omega_{\rm F} = 0.5$ that
the energy contribution of the induced wave is negligible and then 
$\sqrt{-g} \delta^2 {S}^t$
is almost identical to $\sqrt{-g} \delta^{1+1} {S}^t
\equiv \sqrt{-g} \delta^{1+1} {S}^t$.
In fact, the energy of the Alfv\'{e}n wave is almost the same as the total energy
(Fig. \ref{fig_ow_his5t_bzof}), and
the energy density of the Alfv\'{e}n wave remains positive since the initial stage.

\begin{figure} % [H]
\begin{center}
\includegraphics[width=17cm]{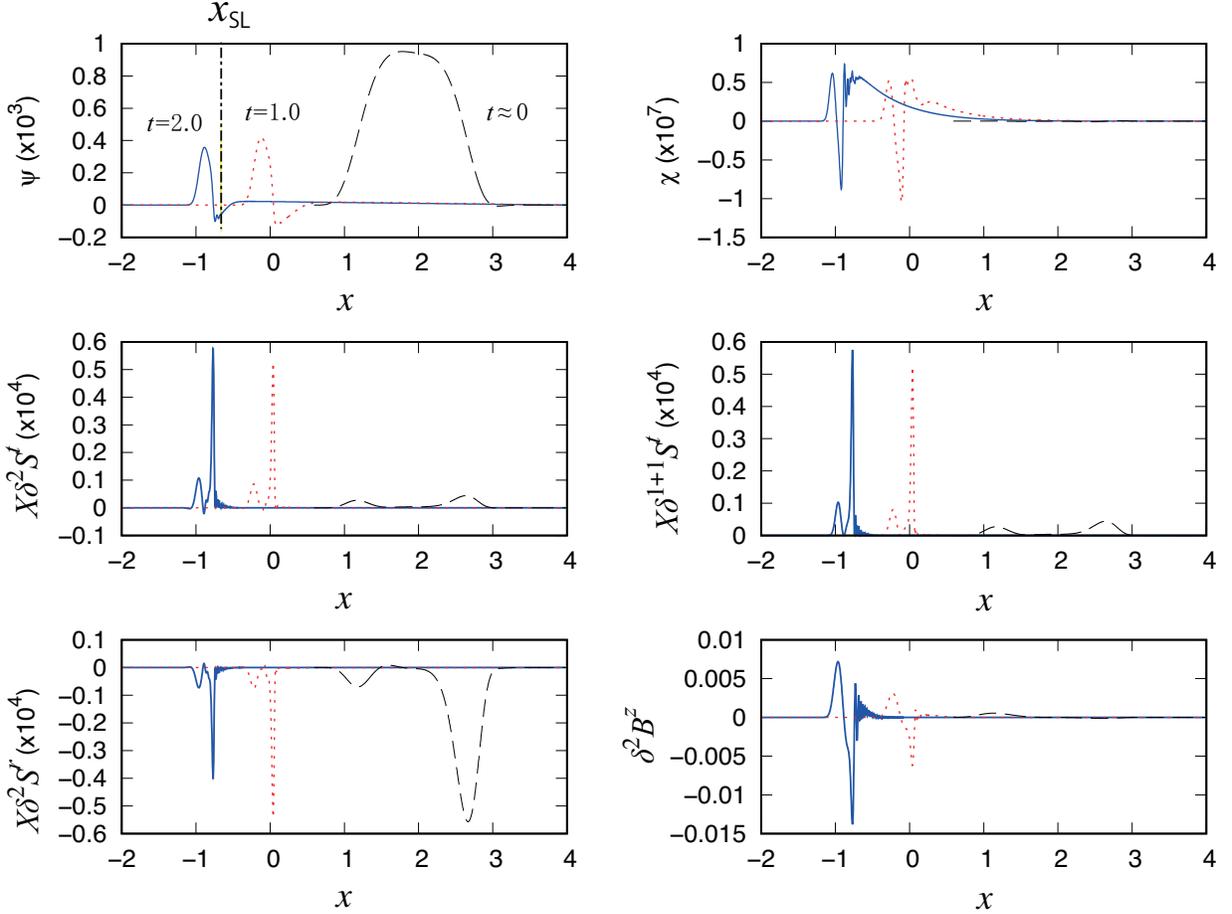}% Here is how to import EPS art
\end{center}
\caption{Similar to Fig. \ref{fig_ow_times}, but for the calculation
of the Alfv\'{e}n wave propagating inwardly from the outer region
at $t=$ 0.01 (black dashed line), 1.0 (red dotted line), 2.0 (blue, thick solid line).
\label{fig_iw_times}}
\end{figure}

\begin{figure} % [H]
\begin{center}
\includegraphics[width=17cm]{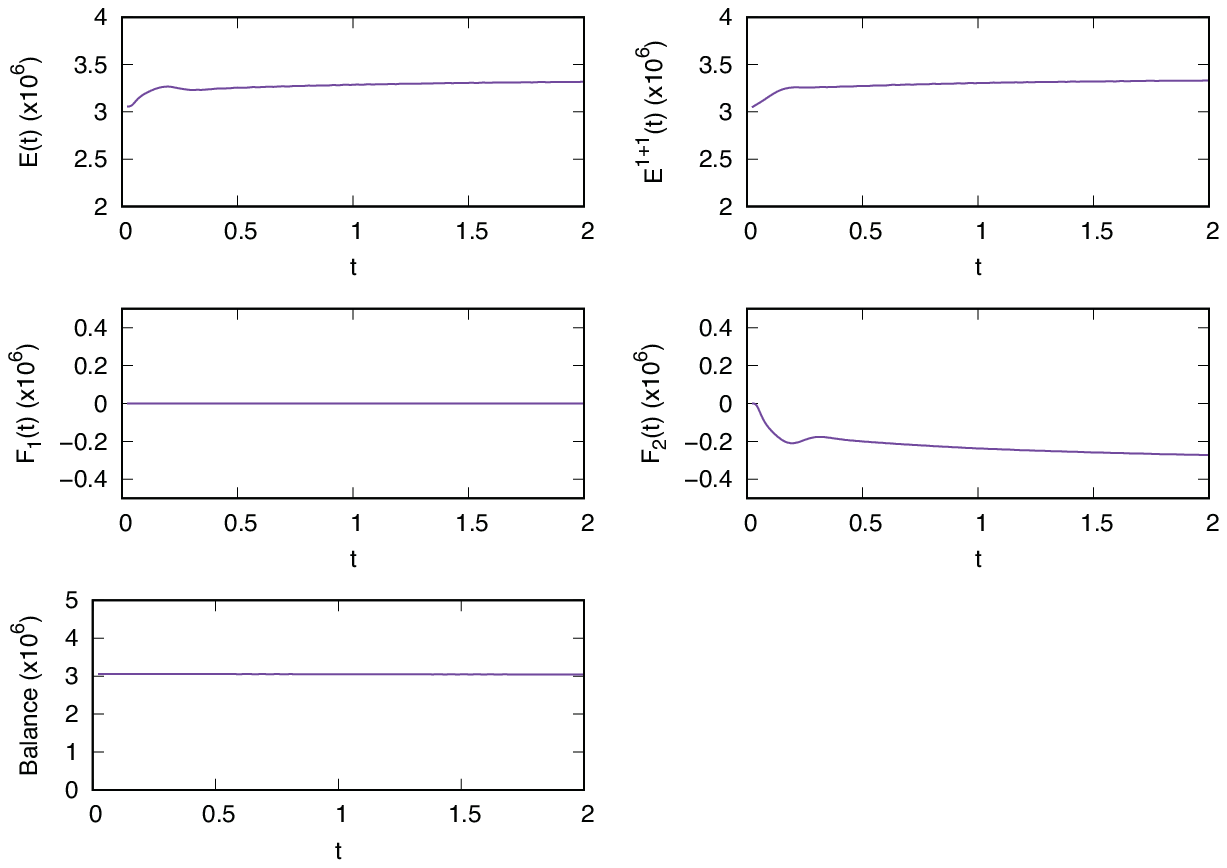}% Here is how to import EPS art
\end{center}
\caption{Similar to Fig. \ref{fig_ow_his5t}, but for the calculation 
of the Alfv\'{e}n wave propagating inwardly from the outer region.
\label{fig_iw_his5t}}
\end{figure}

Figure \ref{fig_iw_times} shows the numerical result of the inwardly propagating
pulse of the 
Alfv\'{e}n wave along the magnetic field line of $\Omega_{\rm F}=0.5$ around
the spinning black string with $a=0.9$.
The Alfv\'{e}n wave pulse propagates inwardly and induces the fast wave.
The induced wave propagates mainly inwardly with the Alfv\'{e}n wave pulse.
The energy contribution of the induced wave is not so large
that $\sqrt{-g} \delta^2 {S}^t$ is almost equal to 
$\sqrt{-g} \delta^{1+1} {S}^t$.
In fact, the time evolution of the total energy almost identical to
that of the Alfv\'{e}n wave (Fig. \ref{fig_iw_his5t}). 
Then, in the case of the inwardly propagating wave from the outer region,
the fast induced wave is negligible for the energy conservation law.
In conclusion, the fast wave induced by the Alfv\'{e}n wave
has a significant role in the case of outwardly propagating Alfv\'{e}n wave
along the rotating curved magnetic field line 
with $0 < \Omega_{\rm F} < \Omega_{\rm H}$.
% where the Blandford--Znajek mechanism works.
However, strictly speaking, we have to consider the energy contribution
of the induced fast wave by the Alfv\'{e}n wave
to guarantee the energy conservation and the force-free condition in the case 
of the Alfv\'{e}n wave propagating
along the rotating curved magnetic field line in the frame work of
relativity. 
%% The effect is negligible on in the non-relativistic (Newtonian) frame work.
In a non-relativistic frame work, the angular momentum (momentum) of the
Alfv\'{e}n wave vanishes and such fast wave is never induced.

\section{Concluding remarks
\label{secdis}}

We performed FFMD numerical simulations of Alfv\'{e}n waves along a magnetic
field line around a black string using Eq. (\ref{nodaeq16dd}) of $\psi$ 
and Eq. (\ref{chieq}) of $\chi$,
which recovers the force-free condition and energy conservation law in the BTZ 
coordinates. In the case of outwardly propagating Alfv\'{e}n waves along the
rotating curved magnetic field line
with $0 < \Omega_{\rm F} < \Omega_{\rm H}$,
% , when the Blandford--Znajek mechanism works, 
the energy of the Alfv\'{e}n wave is initially negative in the ergosphere
and increases to be a positive value out of the ergosphere. 
The Alfv\'{e}n wave induces a second-order fast wave along the magnetic field line.
The total energy of the Alfv\'{e}n and induced fast waves is conserved, and 
$\chi$, which expresses the induced fast wave, plays a significant role in 
energy transport in the BTZ coordinates.

\begin{figure} % [H]
\begin{center}
\includegraphics[width=9cm]{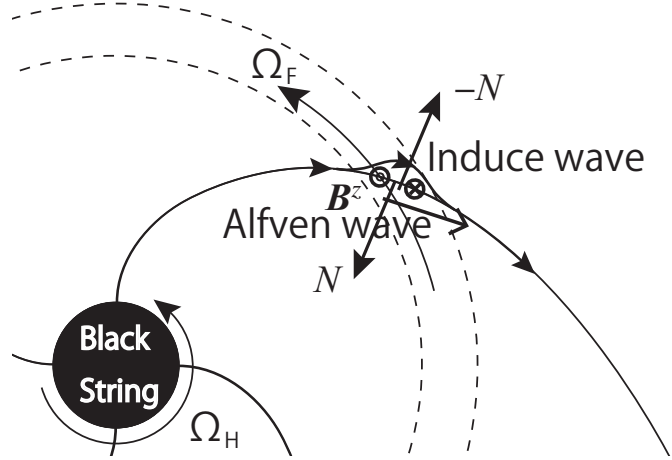}% Here is how to import EPS art
\end{center}
\caption{Schematic of the wave that was induced by 
Alfv\'{e}n wave propagation along 
curved rotating field lines around a spinning black string.
$N$ shows the force from the background magnetic-field line
to the Alfv\'{e}n wave. The reaction of the force produces the induced wave
oscillating in the azimuthal direction on the magnetic field line.
\label{pontie_indwav}}
\end{figure}

The fast wave induced by the Alfv\'{e}n wave is attributed to
the angular momentum of the Alfv\'{e}n wave (Fig. \ref{pontie_indwav}) \footnote{
In the special relativistic framework, 
the momentum density of the transverse wave is given by 
$\vec{S}/c$, where $\vec{S}$ is the
Poynting flux and $c$ is the speed of light.
In the nonrelativistic limit ($c \longrightarrow \infty$), 
the momentum density vanishes.}.
When the Alfv\'{e}n wave propagates along a curved
magnetic field line, the angular momentum changes due to 
the torque from the background magnetic field.
The reaction of the torque induces the fast wave described by $\chi$. 
The above mechanism is expected to work in the relativistic 
magnetohydrodynamic (MHD) framework.
Notably, in non-relativistic MHD,
the angular momentum of the Alfv\'{e}n wave vanishes, and such a fast wave is not induced.
The second-order fast wave would be induced by the Alfv\'{e}n wave
in relativistic MHD.

As shown in Fig. \ref{fig_ow_times}, 
with the energy extraction
from the black string by the Blandford--Znajek
mechanism ($0 < \Omega_{\rm F} < \Omega_{\rm H}$),
the energy of the Alfv\'{e}n wave around the ergosphere is negative,
and without energy extraction from the black string 
by the Blandford--Znajek mechanism
($\Omega_{\rm F} > \Omega_{\rm H}$), the energy is positive all the time
(Fig. \ref{fig_ow_plt6t_bzof}).
The difference between the two cases is explained
by the following inequality derived from Eqs. (\ref{d2sx}) and (\ref{d11st}) with $\chi=0$,
\begin{eqnarray}
\delta^{1+1} S^t 
&=& \frac{\Gamma^2-I'^2 \Gamma'}{2 \alpha^2 X^2 \Gamma^2} (\partial_T \psi)^2
+ \frac{-\Gamma'}{2 X^2} (\partial_X \psi)^2
- \frac{I' X}{\alpha^2 \Gamma^2}
\left (2 \alpha^2 \Omega_{\rm F} + \Gamma \Omega \right ) \delta^{1+1} \underline{S}^X ,
 \nonumber 
% \\
% &>& - \frac{2 I' \Omega_{\rm F} X}{\Gamma^2} \delta^2 S^X
% \left (1 + \frac{\Gamma \Omega}{2 \alpha^2 \Omega_{\rm F}} \right )
% \delta^{1+1} S^X,
\label{d11stineq}
\end{eqnarray}
where $\Gamma'=-\alpha^2 + X^2 (\Omega^2 - \Omega_{\rm F}^2)
= \Gamma + 2 \Omega_{\rm F} W = (1+\Omega_{\rm F}^2) X^2 -1$.
Here, we consider the condition of the negative energy density ($\delta^{1+1} S^t < 0$)
of the Alfv\'{e}n wave propagating outwardly ($\delta^{1+1} \underline{S}^X > 0$).
With respect to the zero point of $\Gamma'$, $X_{\rm c} = 1/(1+\Omega_{\rm F}^2)$,
we have
\begin{equation}
X_{\rm c} - X_{\rm LS} = - \frac{a}{1-\Omega_{\rm F}^4} 
(\Omega_{\rm F}^+ -\Omega_{\rm F}) \Omega_{\rm F} (\Omega_{\rm F}-\Omega_{\rm H}),
\end{equation}
where $X_{\rm LS}=r_{\rm LS}$ and $\Omega_{\rm F}^+ \equiv (1 + \sqrt{1-a^2})/a > 1
> \Omega_{\rm F}$.
Then, if $\Omega_{\rm F} (\Omega_{\rm F}-\Omega_{\rm H}) > 0$,
we find $X_{\rm c} < X_{\rm LS}$ and we have $\Gamma' < 0$.
If $\Gamma' < 0$, the first and second terms of the left-hand side
of Eq. (\ref{d11stineq}) are positive.
When $\Omega_{\rm F} < 0$, we have $\Omega_{\rm F} (\Omega_{\rm F}-\Omega_{\rm H}) > 0$,
$I'=r_{\rm H} (\Omega_{\rm H}-\Omega_{\rm F})>0$
and $2 \alpha^2 \Omega_{\rm F} + \Gamma \Omega < 0$. 
Then, Eq. (\ref{d11stineq}) shows that if $\Omega_{\rm F} < 0$,
$\delta^{1+1} S^t$ is positive because of $\delta^{1+1} \underline{S}^X > 0$.
When $\Omega_{\rm F} > \Omega_{\rm H}$, we also have 
$\Omega_{\rm F} (\Omega_{\rm F}-\Omega_{\rm H}) > 0$,
$I'=r_{\rm H} (\Omega_{\rm H}-\Omega_{\rm F})<0$,
and $2 \alpha^2 \Omega_{\rm F} + \Gamma \Omega =
\Gamma W + \Omega_{\rm F} (\alpha^2 + X^2 W^2) > 0$
because of $W = \Omega - \Omega_{\rm F} < \Omega_{\rm H} - \Omega_{\rm F}<0$. 
Then, Eq. (\ref{d11stineq}) shows that if $\Omega_{\rm F} > \Omega_{\rm H}$,
$\delta^{1+1} S^t$ is positive.
In summary, if $\Omega_{\rm F} < 0$ or $\Omega_{\rm F} > \Omega_{\rm H}$,
$\delta^{1+1} S^t$ is positive when $\delta^{1+1} \underline{S}^X > 0$.
% Eventually, we conclude that when we 
Considering the outwardly propagating Alfv\'{e}n wave ($\delta^{1+1} \underline{S}^X > 0$),
to realize $\delta^{1+1} S^t < 0$, we require $0 < \Omega_{\rm F} < \Omega_{\rm H}$.
This condition is the same as that of the black string rotational energy
extraction due to the Blandford--Znajek mechanism.
% If the energy of the outwardly propagating Alfv\'{e}n wave is negative
% ($\delta^{1+1} S^t < 0$), Eq. (\ref{d11stineq}) indicates that
% $\Omega_{\rm F} I' = r_{\rm H} \Omega_{\rm F} (\Omega_{\rm H} - \Omega_{\rm F})$
% should be positive, that is, $0 < \Omega_{\rm F} <  \Omega_{\rm H}$.
% Here, the factor of the right-hand side of Eq.  (\ref{d11stineq}),
% $\displaystyle 1 + \frac{\Gamma \Omega}{2 \alpha^2 \Omega_{\rm F}}$,
% is positive for $\Omega_{\rm F} =0.5$ and $0.7$.
% This condition is identical with the condition of the Blandford--Znajek mechanism on.
% % This identity between the negative energy of the Alfv\'{e}n wave and
% % the Blandford--Znajek mechanism has no physical meaning,
% % such as physical relationship between the Alfv\'{e}n wave
% % and the Blandford--Znajek mechanism.
Notably, $ 0 < \Omega_{\rm F} < \Omega_{\rm H}$ is the necessary
condition for the negative energy of the Alfv\'{e}n wave, 
which is the necessary and sufficient condition for energy extraction
from the black string by the Blandford--Znajek mechanism.

\begin{figure} % [H]
\begin{center}
\includegraphics[width=10cm]{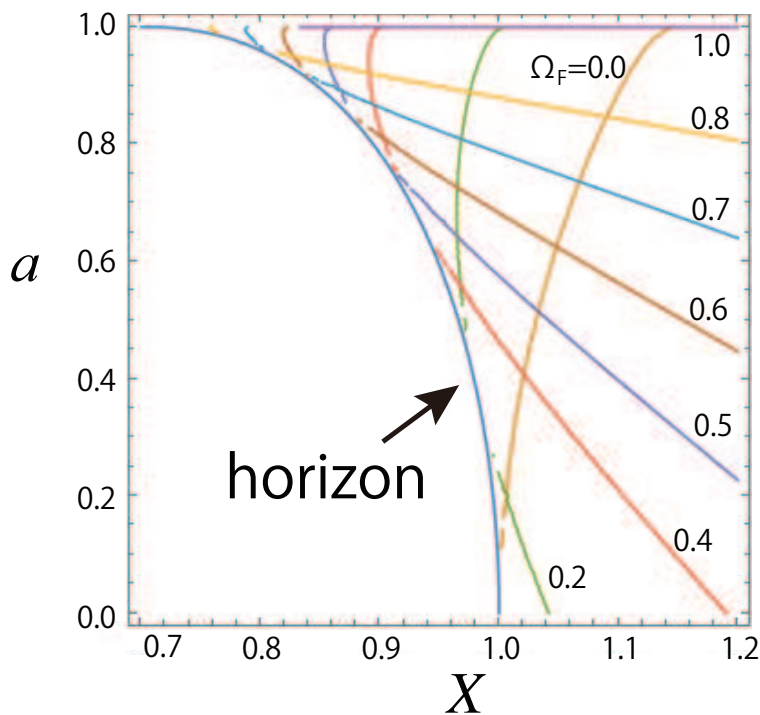}% Here is how to import EPS art
\end{center}
\caption{Position of the horizon and apparent singularity of the
natural coordinates for each black string spin parameter $a$.
\label{fig_snglry_nat}}
\end{figure}

To calculate the force-free field of the Alfv\'{e}n and 
induced fast waves,
we use Eq. (\ref{nodaeq16dd}) of $\psi$ and 
Eq. (\ref{chieq}) of $\chi$ on the natural coordinates
$x^\mu=(T, X, \rho, z)$. There is an apparent singularity
point on the natural coordinates, 
such as $\displaystyle g^{TT}=-\frac{\gamma_{\rm N}}{X}=0$.
The position of the apparent singularity of the natural coordinates
is shown in Fig. \ref{fig_snglry_nat}.
Eq. (\ref{nodaeq16dd}) does not contain $g^{TT}$, and then, we calculate the Alfv\'{e}n
wave over the region $X>X_{\rm LS}$. However, the time-evolution equation
of $\chi$ contains $g^{TT}$ explicitly. 
% We have to avoid the calculation of 
Thus, $\chi$ is not to be calculated
on the singular point.
The natural coordinates ease the calculations of the equations drastically,
but it has a restriction of the numerical calculation region
with respect to the singularity. We would solve the
problem of the apparent singularity with the natural coordinates
in our next report.

\begin{acknowledgments}
We are grateful to Mika Koide for her helpful comments on this paper.
M.T. was supported in part by JSPS KAKENHI Grant No. 17K05439.
\end{acknowledgments}



\appendix

\section{Energy-momentum tensor and 
force-free condition for Alfv\'{e}n wave
\label{appenda}}
We show the detailed calculation with respect to the energy-momentum tensor and 
the force-free condition in the case
of the Alfv\'{e}n wave only, where we assume
\begin{equation}
\phi_1 = \bar{\phi}_1 + \delta \phi_1 = -z + \psi, \verb!   !
\phi_2 = \bar{\phi}_2.
\end{equation}
First, we show non-zero components of the field tensor of the equilibrium, 
first and second-order perturbations as follows.
% we have the following non-vanishing components of the field tensor.
Non-vanishing components of the field tensor of the equilibrium are only
\begin{equation}
\bar{F}_{\rho z} = - \bar{F}_{z \rho} =1, \verb!   !
\bar{F}^{\mu z} = - \bar{F}^{z \mu} = g^{\mu \rho} \verb!   ! (\mu \ne z).
\end{equation}
Non-zero components of the field tensor of the first-order perturbations are only
\begin{equation}
\delta F_{\lambda \rho}  =  - \delta F_{\rho\lambda} = \partial_{\lambda} \psi
\verb!   ! (\lambda \ne \rho) , \verb!   !
\delta F^{\mu \nu}  =   W^{\mu\lambda\nu\rho} \partial_\lambda \psi.
\end{equation}
All components of the field tensor of the second-order perturbations vanish.\\
% \begin{equation}
% \delta^2 F_{\mu z}  =  - \delta^2 F_{z \mu} = \partial_{\mu} \chi 
% \verb!    ! (\mu \ne z) , \verb!   !
% \delta^2 F^{\mu \nu}  =   - W^{\mu z\nu\beta} \partial_\beta \chi.
% \end{equation}
Then, we calculate
\begin{eqnarray}
\bar{F}^{\mu\lambda} \delta F_{\nu \lambda} 
&=& \bar{F}^{\mu z} \delta F_{\nu z} = g^{\mu \rho} \delta F_{\nu z} = 0, \\
\bar{F}^{\lambda\kappa} \delta F_{\lambda\kappa} 
&=& \bar{F}_{\lambda\kappa} \delta F^{\lambda\kappa}
= 2 \bar{F}_{\rho z} \delta F^{\rho z} = 2 W^{\rho\lambda z \rho} \partial_\lambda \psi
= -2 g^{\rho\rho} g^{\lambda z} \partial_\lambda \psi=0.
\end{eqnarray}
We find
\begin{eqnarray}
\delta F^{\mu\lambda} \bar{F}_{\nu\lambda}
&=& \delta F^{\mu z} \bar{F}_{\nu z} + \delta F^{\mu\rho} \bar{F}_{\nu\rho}
= \delta F^{\nu z} \delta_{\nu\rho} - \delta^{\mu\rho} \delta_{\nu z}
= W^{\mu\lambda z\rho} \partial_{\lambda} \psi \delta_{\nu\rho}
- W^{\mu\lambda \rho\rho} \partial_{\lambda} \psi \delta_{\nu z} \nonumber \\
&=& - g^{\mu\rho} g^{zz} \partial_z \psi \delta_{\nu\rho}
- W^{\mu\lambda\rho\rho} \partial_\lambda \psi \delta_{\nu z}
= -W^{\mu\lambda\rho\rho} \partial_\lambda \psi \delta_{\nu z}=0.
\end{eqnarray}
Because when $\nu = \rho$, we have $\delta F_{\rho z} = - \partial_z \psi = 0$,
and otherwise, we have $\delta F_{\nu z}=0$.
Then, we conclude that when $\mu$ and $\nu$ are not $z$, we have
\begin{equation}
\delta T^\mu_\nu = 0.
\end{equation}

To check the force-free condition, we calculate the Lorentz force,
\begin{equation}
f_\mu^{\rm L} = J^\nu F_{\mu\nu}, \verb!   !
J^\mu = \nabla_\nu F^{\mu\nu} = \frac{1}{\sqrt{-g}} \partial_\nu (\sqrt{-g} F^{\mu\nu}).
\end{equation}
First, we can confirm the Lorentz force of the equilibrium vanishes because
the 4-current density of the equilibrium vanishes:
\begin{equation}
\bar{J}^\mu = \frac{1}{\sqrt{-g}} \partial_\nu (\sqrt{-g} \bar{F}^{\mu\nu})
= \frac{1}{\sqrt{-g}} \partial_X (\sqrt{-g} \bar{F}^{\mu X})=0,
\verb!   !
\bar{f}_\mu^{\rm L} = \bar{J}^\nu \bar{F}_{\mu\nu}=0,
\end{equation}

Second, we confirm the Lorentz force with respect to the first-order perturbation
of Alfv\'{e}n wave vanishes:
\begin{eqnarray}
\delta f_{i}^{\rm L} &=& \delta J^\nu \bar{F}_{i \nu} = 0, \\
\delta f_{\rho}^{\rm L} &=& \delta J^\nu \bar{F}_{\rho\nu} 
= \delta J^z \bar{F}_{\rho z} = \delta J^z
= \frac{1}{\sqrt{-g}} \partial_\nu (\sqrt{-g} \bar{F}^{z \nu})
= \frac{1}{\sqrt{-g}} \partial_\nu (\sqrt{-g} W^{z \lambda\nu\rho} \partial_\lambda \psi)
= 0, \\
\delta f_{z}^{\rm L} &=& \delta J^\nu \bar{F}_{z \nu} 
= \delta J^\rho \bar{F}_{z \rho} = - \delta J^\rho
= - \frac{1}{\sqrt{-g}} \partial_\nu (\sqrt{-g} \delta F^{\rho \nu}) \nonumber \\
& = & -\frac{1}{\sqrt{-g}} \partial_\nu (\sqrt{-g} W^{\rho\lambda\nu\rho} \partial_\lambda \psi)
= \frac{1}{\sqrt{-g}} \partial_\nu (\sqrt{-g} W^{\nu\lambda\rho\rho} \partial_\lambda \psi)
= 0,
\label{forcefreecheck}
\end{eqnarray}
where $i$ is $T$ or $X$.
In the last equation of Eq. (\ref{forcefreecheck}), we use Eq.
(\ref{nodaeq16dd}).
Then, we confirmed the first order of the Lorentz force vanishes.

It is noted that we have $\delta J^\rho = \delta J^z = 0$.
%%% To obtain a guess, first we calculate the Lorentz force of the simplest case:
%%% \begin{equation}
%%% \delta^2 \phi_1 = 0, \verb!   ! \delta \phi_2 =  \delta^2 \phi_2 =0.
%%% \end{equation}
Using
\begin{equation}
\delta J^\mu = \frac{1}{\sqrt{-g}} \partial_\nu (\sqrt{-g} \delta F^{\mu\nu})
=  \frac{1}{\sqrt{-g}} \partial_\nu (\sqrt{-g} W^{\mu\lambda\nu\rho} \partial_\lambda \psi),
\end{equation}
we have 
\begin{eqnarray}
\delta J^T & =& \frac{1}{\sqrt{-g}} \sum_{\lambda = T, X} 
\partial_\nu (\sqrt{-g} W^{T \lambda\nu\rho} \partial_\lambda \psi)
= \frac{1}{\sqrt{-g}} \sum_{\lambda = T, X} 
\partial_X (\sqrt{-g} W^{T \lambda X\rho} \partial_\lambda \psi) ,\\
\delta J^X & =& \frac{1}{\sqrt{-g}} \sum_{\lambda = T, X} 
\partial_\nu (\sqrt{-g} W^{X \lambda\nu\rho} \partial_\lambda \psi)
= - \frac{1}{\sqrt{-g}} \sum_{\lambda = T, X} 
\partial_X (\sqrt{-g} W^{T \lambda X \rho} \partial_\lambda \psi), \\
\delta J^\rho &=& \delta J^z = 0,
\end{eqnarray}
where $\displaystyle W^{TTX\rho} = \frac{I'}{X \Gamma}$ and 
$W^{TXX\rho} = W = \Omega - \Omega_{\rm F}$.
We also have
\begin{equation}
\delta^2 J^\mu = \frac{1}{\sqrt{-g}} \partial_\nu (\sqrt{-g} \delta^2 F^{\mu\nu})=0 .
\end{equation}

We find three components of the  Lorentz force vanish as
\begin{eqnarray}
\delta^2 f_{T}^{\rm L} &=& \delta J^\mu \delta F_{T \mu} 
= \delta J^\rho \delta F_{T \rho} = 0,\\
\delta^2 f_{X}^{\rm L} &=& \delta J^\mu \delta F_{X \mu} 
= \delta J^\rho \delta F_{X \rho} = 0,\\
\delta^2 f_{z}^{\rm L} &=& \delta J^\mu \delta F_{z \mu} 
= \delta J^\rho \delta F_{z \rho} = 0 ,
\end{eqnarray}
because of $\delta J^\rho = 0$.
The $\rho$-component of the Lorenz force is calculated as,
\begin{eqnarray}
\delta^2 f_{\rho}^{\rm L} &=& \delta J^\mu \delta F_{\rho\mu} 
= \delta J^T \delta F_{\rho T} + \delta J^X \delta F_{\rho X}
= - \delta J^T \partial_T \psi - \delta J^X \partial_X \psi \nonumber \\
&=& \frac{1}{\sqrt{-g}} \sum_{\lambda = T, X} 
[ - \partial_X (\sqrt{-g} W^{T \lambda X \rho} \partial_\lambda \psi) \partial_T \psi
+ \partial_T (\sqrt{-g} W^{T \lambda X \rho} \partial_\lambda \psi) ] \partial_X \psi \nonumber \\
&=& \frac{1}{\sqrt{-g}}
[ - \partial_X \{ \sqrt{-g} \partial_T \psi (W^{TTX\rho} \partial_T \psi
+W^{TXX\rho} \partial_X \psi) \}
 + \partial_T \{ \sqrt{-g} \partial_X \psi (W^{TTX\rho} \partial_T \psi
+W^{TXX\rho} \partial_X \psi) \} ] \nonumber \\
&=& \frac{1}{\sqrt{-g}}
\left [ - \partial_X \left \{ \partial_T \psi \left ( \frac{I'}{\Gamma} \partial_T \psi
+ X W \partial_X \psi \right ) \right \}
 + \partial_T \left \{ \partial_X \psi \left (\frac{I'}{\Gamma} \partial_T \psi
+ X W \partial_X \psi \right) \right \} \right ] .
\end{eqnarray}
%%% $\delta^2 f_{\rho}^{\rm L}$ vanishes only if $I'=0$ and $W= \Omega - \Omega_{\rm F}=0$.
%%% Then, when $I'$ and $W$ are finite, $\delta^2 f_{\rho}^{\rm L}$ is finite and
%%% the force-free condition is broken.
%%% In fact, as shown by the numerical simulations in Section \ref{secmet},
%%% the conservation of energy calculated only with $\psi$ is broken.

\section{Calculation of energy density and energy flux density
corrected by $\chi$
\label{appendb}}

We show the detail of the calculation of the energy density and energy flux
of Alfv\'{e}n wave and the induced fast wave described by $\chi$.
% given by $\psi$ and $\chi$, respectively.
% To recover the force-free condition, we should take into account
% of the additional first and second-order perturbations $\delta^2 \phi_1$, 
% $\delta \phi_2$, $\delta^2 \phi_2$ as
% \begin{eqnarray}
% \phi_1 &= \bar{\phi}_1 + \delta \phi_1 + \delta^2 \phi_1 = -z + \psi + \delta^2 \phi_1, \\
% \phi_2 &= \bar{\phi}_2 + \delta \phi_2 + \delta^2 \phi_2 
% = \rho + \delta \phi_2 + \delta^2 \phi_2.
% \end{eqnarray}
% We should notice that the force-free condition is broken only on one
% equation with respect to $\delta^2 J_\rho$. Then, the breakdown of 
% the force-free condition should be recovered by addition of 
% one freedom variable. The above intuitive reason of the Lorentz force acting on
% the Alfv\'{e}n wave suggests the variable $\delta^2 \phi_2$ is the appropriate
% additional perturbation as the additional freedom.
% To recover the force-free condition 
% we consider the following perturbation setting of the force-free Alfv\'{e}n wave
% along the magnetic field line around the spinning black string:
The Alfv\'{e}n wave and the fast wave are given by
the following perturbation, respectively,
\begin{eqnarray}
\phi_1 &=& \bar{\phi}_1 + \delta \phi_1 = -z + \psi(T, X), \\
\phi_2 &=& \bar{\phi}_2 + \delta^2 \phi_2 = \rho + \chi(T, X) .
\end{eqnarray}
%%% \subsection{Equation of $\chi$}
To derive time evolution equation of $\chi$, we use Eq. (\ref{nodaeq04w}).
When we take $i=2$, we find the trivial equation:
\begin{equation}
\partial_\lambda (\sqrt{-g} W^{\lambda\alpha\mu\beta} \partial_\alpha \bar{\phi}_1
\partial_\beta \chi ) \partial_\mu \bar{\phi}_2
+ \partial_\lambda (\sqrt{-g} W^{\lambda\alpha\mu\beta} \partial_\alpha \bar{\phi}_1
\partial_\beta \bar{\phi}_2 ) \partial_\mu \chi
= - \partial_\lambda (\sqrt{-g} W^{\lambda z \rho\beta} \partial_\beta \chi)
- \partial_\lambda (\sqrt{-g} W^{\lambda z \mu\rho} \partial_\mu \chi) = 0.
\end{equation}
When we take $i=1$, we obtain the equation of $\chi$:
\begin{eqnarray}
&& \partial_\lambda (\sqrt{-g} W^{\lambda\alpha\mu\beta} \partial_\alpha \psi
\partial_\beta \bar{\phi}_2 ) \partial_\mu \psi
+ \partial_\lambda (\sqrt{-g} W^{\lambda\alpha\mu\beta} \partial_\alpha \bar{\phi}_1
\partial_\beta \chi) \partial_\mu \bar{\phi}_1     \nonumber \\
&=& \partial_\lambda (\sqrt{-g} W^{\lambda\alpha\mu\rho} \partial_\alpha \psi) 
\partial_\mu \psi
+ \partial_\lambda (\sqrt{-g} W^{\lambda zz \beta} \partial_\beta \chi)  \nonumber \\
&=& \partial_\lambda (-\sqrt{-g} g^{\lambda\beta} \partial_\beta \psi) 
+ \partial_\lambda (\sqrt{-g} W^{\lambda\alpha\mu\rho} \partial_\alpha \psi) 
\partial_\mu \psi = 0.
\end{eqnarray}
Then, we obtain the time evolution equation of $\chi$,
\begin{equation}
\Box \chi = \nabla_\mu \nabla^\mu \chi = \frac{1}{\sqrt{-g}} \partial_\lambda
(\sqrt{-g} g^{\mu\beta} \partial_\beta \chi)
= \frac{1}{\sqrt{-g}} \partial_\lambda (\sqrt{-g} W^{\lambda\alpha\mu\rho}
\partial_\alpha \psi \partial_\mu \psi) \equiv s.
\label{eqchi4dlb_ap}
\end{equation}
% Eq. (\ref{eqchi4dlb_ap}) reads
% \begin{equation}
% \Box \chi = \frac{1}{\sqrt{-g}} \left [
% \sqrt{-g} g^{TT} \partial_T^2 \chi + 2 f \partial_x ( \sqrt{-g} g^{TX} \partial_T \chi )
% - \partial_x(\sqrt{-g}) \partial_T \chi + f \partial_x (\sqrt{-g} g^{XX} f \partial_x \chi)
% \right ] = s.
% \label{chieq_ap}
% \end{equation}
When we consider $\chi$, we found the force-free condition recovers as follows.
When we take $\nu \ne \rho$, we have
\begin{equation}
\delta^2 f_\nu = \delta J^\mu \delta F_{\nu \mu} + \delta^2 J^\mu \bar{F}_{\nu \mu}
= \delta J^\rho \delta F_{\nu \rho} + \delta^2 J^\rho \bar{F}_{\nu \rho} = 0. 
\end{equation}
Otherwise, we have
\begin{align}
\delta^2 f_\rho &= \delta J^\mu \delta F_{\rho \mu} + \delta^2 J^\mu \bar{F}_{\rho \mu}
= - \frac{1}{\sqrt{-g}} \partial_\nu (\sqrt{-g} W^{\mu\lambda\nu\rho} \partial_\lambda \psi)
\partial_\mu \psi + \delta^2 J^z  \nonumber \\
& = - \frac{1}{\sqrt{-g}} \partial_\nu (\sqrt{-g} g^{\nu\beta} \partial_\beta \chi)
+ \frac{1}{\sqrt{-g}} \partial_\nu (\sqrt{-g} W^{\nu\lambda\mu\rho} \partial_\lambda \psi
\partial_\mu \psi) = 0.
\end{align}
Eventually, introducing $\chi$, we recovered and confirmed the force-free condition.

%%% \subsection{Energy and angular momentum conservation law 
%%% in the corotating natural coordinates}

% In the corotating natural coordinates, $f_T^{\rm L}$ vanishes, then
% the energy conservation law stands even without introducing $\chi$.
% However, including $\chi$, we have to recalculate the energy density
% and energy flux.
% The equilibrium and first order of these values are the same
% as the calculation without introducing $\chi$.
% % % To evaluate the perturbations,
% % % we show non-zero components of the field tensor of the equilibrium, 
% % % first and second-order perturbations as follows.
We show the detail derivation of the equilibrium, first order, and second order
of the values with respect to the conservation law.
% % %
% First, we show non-zero components of the field tensor of the equilibrium, 
% first and second-order perturbations as follows.
% Non-vanishing components of the field tensor of the equilibrium are only
% \begin{eqnarray}
% \bar{F}_{\rho z} & = & - \bar{F}_{z \rho} =1, \\
% \bar{F}^{\mu z} & = & - \bar{F}^{z \mu} = g^{\mu \rho} \verb!   ! (\mu \ne z).
% \end{eqnarray}
% Non-zero components of the field tensor of the first-order perturbations are only
% \begin{eqnarray}
% \delta F_{\lambda \rho} & = & - \delta F_{\rho\lambda} = \partial_{\lambda} \psi
% \verb!   ! (\lambda \ne \rho) , \\
% \delta F^{\mu \nu} & = &  W^{\mu\lambda\nu\rho} \partial_\lambda \psi.
% \end{eqnarray}
% Non-vanishing components of the field tensor of the second-order perturbations are only
% \begin{eqnarray}
% \delta^2 F_{\mu z} & = & - \delta^2 F_{z \mu} = \partial_{\mu} \chi 
% \verb!    ! (\mu \ne z) , \\
% \delta^2 F^{\mu \nu} & = &  - W^{\mu z\nu\beta} \partial_\beta \chi.
% \end{eqnarray}
With respect to the energy density, we have
\begin{eqnarray}
\bar{S}^0 &=& -\frac{1}{2} \bar{F}^{0i} \bar{F}_{0i} 
+ \frac{1}{4} \bar{F}^{ij} \bar{F}_{ij} = \frac{1}{2} g^{\rho\rho}
= \frac{I'^2 - \Gamma}{2 X^2 {\alpha}^2} , \\
\delta S^0 &=&  -\frac{1}{2} \delta F^{0i} \bar{F}_{0i} 
- \frac{1}{2} \bar{F}^{0i} \delta F_{0i}
+ \frac{1}{4} \delta F^{ij} \bar{F}_{ij}+ \frac{1}{4} \bar{F}^{ij} \delta F_{ij} 
= -g^{\rho\rho} \partial_z \psi = 0, \\
\delta^2 S^0 &=&  -\frac{1}{2} \delta F^{0i} \delta F_{0i} 
+ \frac{1}{4} \delta F^{ij} \delta F_{ij} 
-\frac{1}{2} \bar{F}^{i0} \delta^2 F_{0i} + \frac{1}{4} \bar{F}^{ij} \delta^2 F_{ij}
+ \frac{1}{4} \delta^2 F^{ij} \bar{F}_{ij}  \nonumber \\
&=& \frac{-1}{2 \Gamma} \left [\frac{-\Gamma+I'^2}{{\alpha}^2 X^2} (\partial_T \psi)^2 
+ \frac{\Gamma^2}{X^2} (\partial_X \psi)^2 \right ] 
+ g^{X\rho} \partial_X \chi \nonumber \\
&=& \frac{-\gamma_{\rm F}}{2 \Gamma} [\lambda (\partial_T \psi)^2 +(\partial_x \psi)^2]
+ \frac{I' \gamma_{\rm F}}{-\Gamma} \partial_x \chi.
\end{eqnarray}
With respect to the energy flux, we have 
% to modify only the second-order expression.
\begin{eqnarray}
\bar{S}^X &=& \bar{F}^{Xi} \bar{F}_{i0} = 0,  \\
\delta S^X &=&  \delta F^{Xi} \bar{F}_{i0} + \bar{F}^{Xi} \delta F_{i0} = 0, \\
% \end{eqnarray}
% \begin{eqnarray}
\delta^2 S^X &=&  \delta F^{Xi} \delta F_{i0} 
+ \bar{F}^{Xi} \delta^2 F_{i0}
=  \delta F^{X\rho} \delta F_{\rho 0}
+\bar{F}^{Xz} \delta^2 F_{z0} \nonumber \\
&=& \frac{\Gamma}{X^2} \partial_T \psi \partial_X \psi
- g^{X\rho} \partial_T \chi
= - \frac{\gamma_{\rm F}}{X} \partial_T \psi \partial_x \psi 
- \frac{I'}{X} \partial_T \chi.
\end{eqnarray}

%% The time-like Killing vector $\xi^\mu_{(T)} = (1, 0, 0, 0)$ yields
%% the 4-energy flux density $S^\mu = - \xi^\nu_{(T)} T^\mu_\nu$, and
%% we obtain the energy conservation law in the corotating natural coordinates,
%% \begin{equation}
%% \nabla_\mu S^\mu = \frac{1}{\sqrt{-g}} \partial_\mu (\sqrt{-g} S^\mu)
%% = \frac{\partial S^0}{\partial T} + \frac{1}{\sqrt{-g}} \frac{\partial}{\partial X}
%% (\sqrt{-g} S^X) = \xi^\mu_{(T)} f_\mu^{\rm L} = 0 .
%% \end{equation}

With respect to the angular momentum, we introduced $\chi$
to recover the force-free condition of the Alfv\'{e}n wave
up to second order of the perturbation so that we become able to consider 
the conservation law of angular momentum up to second order.
The axial Killing vector $\xi^\mu_{(\rho)} = (0, 0, 1, 0)$ yields
the 4-angular momentum flux density $M^\mu = \xi^\nu_{(\rho)} T^\mu_\nu$, and
we obtain the angular momentum conservation law in the corotating natural coordinates,
\begin{equation}
\nabla_\mu M^\mu = \frac{1}{\sqrt{-g}} \partial_\mu (\sqrt{-g} M^\mu)
= \frac{\partial M^0}{\partial T} + \frac{1}{\sqrt{-g}} \frac{\partial}{\partial X}
(\sqrt{-g} M^X) = \xi^\mu_{(\rho)} f_\mu^{\rm L} = f_\rho^{\rm L} = 0 .
\end{equation}
The 4-angular momentum density $M^\mu$
in the corotating natural coordinates calculated as
\begin{equation}
M^\mu = \xi^\nu_{(\rho)} T^{\mu}_{\nu} = T^\mu_\rho = F^{\mu \nu} F_{\rho\nu}.
\end{equation}
Then, we have the 4-angular momentum density of the equilibrium, first and
second-order perturbations of the linear Alfv\'{e}n wave,
\begin{eqnarray}
\bar{M}^\mu &=& \bar{T}^\mu_\rho = \bar{F}^{\mu\lambda} \bar{F}_{\rho\lambda} 
= g^{\mu\rho}, \\
\delta M^\mu &=& 
\delta F^{\mu\nu} \bar{F}_{\rho\nu} + \bar{F}^{\mu\nu} \delta F_{\rho\nu}
= \delta F^{\mu z} \bar{F}_{\rho z} + \bar{F}^{\mu z} \delta F_{\rho z}
= \delta F^{\mu z} - g^{\mu \rho} \partial_z \psi = W^{\mu\lambda z\rho}
\partial_\lambda \psi = 0, \\
\delta^2 M^\mu &=& \delta^2 F^{\mu\nu} \bar{F}_{\rho\nu} 
+ \delta F^{\mu\nu} \delta F_{\rho\nu} + \bar{F}_{\mu\nu} \delta^2 F^{\rho\nu}
= \delta^2 F^{\mu z} \bar{F}_{\rho z} 
+ \bar{F}^{\mu z} \delta^2 F_{\rho z} + \delta F^{\mu \nu} \delta F_{\rho\nu} \nonumber \\
& = & - W^{\mu\lambda\nu\rho} \partial_\lambda \psi \partial_\nu \psi
- W^{\mu zz \beta} \partial_\beta \chi - g^{\mu\rho} \partial_z \chi
= - W^{\mu\lambda\nu\rho} \partial_\lambda \psi \partial_\nu \psi
+ g^{\mu\beta} \partial_\beta \chi,
\end{eqnarray}
where we assume $\mu \ne z$.
Then, we confirm the conservation of the angular momentum for the second 
order of the perturbation as
\begin{equation}
\nabla_\mu \delta^2 M^\mu = \frac{1}{\sqrt{-g}} \partial_\lambda (\sqrt{-g} M^\lambda)
= \frac{1}{\sqrt{-g}} \partial_\lambda [\sqrt{-g} (g^{\lambda\beta} \partial_\beta \chi
-W^{\lambda\alpha\mu\rho} \partial_\alpha \psi \partial_\mu \psi)] = 0,
\end{equation}
where we used Eq. (\ref{nodaeq16dd}).
We have the energy density and energy flux density of the equilibrium, first and
second-order perturbations of the linear Alfv\'{e}n wave distinctively,
\begin{eqnarray}
\bar{M}^0 &=& g^{0 \rho} = - \frac{\Gamma - I'^2}{\alpha^2 \Gamma} (\Omega - \Omega_{\rm F}), \verb!   !
\bar{M}^X = g^{X\rho} = \frac{I'}{X}, \\
\delta M^0 &=& W^{0\lambda z\rho} \partial_\lambda \psi = 0, \verb!   !
\delta M^X = W^{X\lambda z\rho} \partial_\lambda \psi = 0, \\
\delta^2 M^0 &=& 
- W^{0\lambda\nu\rho} \partial_\lambda \psi \partial_\nu \psi 
+ g^{0 \beta} \partial_\beta \chi
= - W^{T\lambda X \rho} \partial_\lambda \psi \partial_X \psi 
+ g^{T \beta} \partial_\beta \chi ,\\
\delta^2 M^X &=& 
- W^{X\lambda\nu\rho} \partial_\lambda \psi \partial_\nu \psi 
+ g^{X \beta} \partial_\beta \chi
= W^{T\lambda X \rho} \partial_\lambda \psi \partial_T \psi 
+ g^{X \beta} \partial_\beta \chi .
\end{eqnarray}
%%% These expansions show the second-order perturbation should be considered
%%% to investigate the energy transport of the linear Alfv\'{e}n wave.
%%% Then, the force-free condition should be satisfied up to second order 
%%% of the perturbation. If the force-free condition of the second order 
%%% perturbation is broken, the energy conservation of the Alfv\'{e}n wave
%%% does not hold.
%%% The conservation of the angular momentum of the Alfv\'{e}n wave is the same with
%%% energy conservation of the Alfv\'{e}n wave.
%%% To hold the conservation of the energy and the angular momentum, we checked
%%% the force-free condition up to the second order of the perturbation
%%% of the Alfv\'{e}n wave.

\section{Derivation of energy density and energy flux
in BTZ spacetime
\label{appendc}}

We show the detail derivation of energy density and energy flux
in BTZ spacetime.
The energy density in the BTZ coordinates is calculated by
\begin{eqnarray}
{S}^t & = & - T^t_t 
= - \frac{\partial t}{\partial x^{\underline{\lambda}}} 
\frac{\partial x^{\underline{\kappa}}}{\partial t}
T^{\underline{\lambda}}_{\underline{\kappa}}
= - \left [ \frac{\partial t}{\partial T} \frac{\partial T}{\partial t} T^T_T
+ \frac{\partial t}{\partial T} \frac{\partial \rho}{\partial t} T^T_\rho
+ \frac{\partial t}{\partial X} \frac{\partial T}{\partial t} T^X_T
+ \frac{\partial t}{\partial X} \frac{\partial \rho}{\partial t} T^X_\rho \right ]
\nonumber \\
& = &  - T^T_T + \Omega_{\rm F} T^T_\rho + I_0 (T^X_T - \Omega_{\rm F} T^X_\rho)
=  \underline{S}^T + \Omega_{\rm F} \underline{M}^T - I_0 (\underline{S}^X + \Omega_{\rm F} \underline{M}^X).
\end{eqnarray}
Then, we have the energy density and energy flux density of the equilibrium, first and
second-order perturbations of the linear Alfv\'{e}n wave,
\begin{eqnarray}
{\bar{S}}^t &=& \bar{\underline{S}}^T + \Omega_{\rm F} \bar{\underline{M}}^T
- I_0 (\bar{\underline{S}}^X + \Omega_{\rm F} \bar{\underline{M}}^X)
= \frac{1}{2} g^{\rho\rho} + \Omega_{\rm F} (g^{T\rho} - I_0 g^{X \rho}), \\
\delta {S}^t &=& \delta {\underline{S}}^T + \Omega_{\rm F} \delta {\underline{M}}^T
- I_0 (\delta {\underline{S}}^X + \Omega_{\rm F} \delta {\underline{M}}^X) = 0 , \\
\delta^2 {S}^t &=& \delta^2 {\underline{S}}^T + \Omega_{\rm F} \delta^2 \underline{M}^T
- I_0 (\delta^2 \underline{S}^X + \Omega_{\rm F} \delta^2 \underline{M}^X) = 
\delta^{1+1} {S}^t +  \delta^{2+0} {S}^t  ,
\end{eqnarray}
where
\begin{eqnarray}
\delta^{1+1} {S}^t 
%% & = & -\frac{1}{2} W^{TT\rho\rho} (\partial_T \psi)^2 
%% +\frac{1}{2} W^{XX\rho\rho} (\partial_X \psi)^2 
%% -  \Omega_{\rm F} W^{T\lambda X \rho} \partial_\lambda \psi \partial_X \psi
%% + I_0 (W^{XX\rho\rho} \partial_X \psi \partial_T \psi + \Omega_{\rm F} 
%% W^{X\nu\rho\rho} \partial_\nu \psi \partial_T \psi) \nonumber \\
& = & \frac{1}{2 \Gamma^2} \left [ \frac{1}{X^2 {\alpha}^2} \left \{ 
\Gamma^2 + I'^2({\alpha}^2 - X^2 (\Omega^2 - \Omega_{\rm F}^2)) \right \} (\partial_T \psi)^2
+ \frac{\Gamma^2}{X^2} \left \{ {\alpha}^2 - X^2 (\Omega^2 - \Omega_{\rm F}^2) \right \} 
(\partial_X \psi)^2 \right . \nonumber \\
& &  \left . - 2 \frac{\Gamma}{{\alpha}^2 X} I'
\{ \Gamma (\Omega - \Omega_{\rm F}) + \Omega_{\rm F}
({\alpha}^2 + X^2 (\Omega - \Omega_{\rm F})^2 ) \} \partial_T \psi \partial_X \psi  \right ]
 , \\
\delta^{2+0} {S}^t & = & g^{X \rho} \partial_X \chi + \Omega_{\rm F} 
g^{T \beta} \partial_\beta \chi + I_0 (g^{X \rho} \partial_T \chi 
- \Omega_{\rm F} g^{X \beta} \partial_\beta \chi) , \nonumber \\
& = & \frac{I'}{X} \partial_X \chi - \frac{1}{{\alpha}^2 \Gamma}
(\Omega_{\rm F} \Gamma - I'^2 W) \partial_T \chi.
\end{eqnarray}
When we use the tortoise coordinate $x$, we have the following expressions,
\begin{eqnarray}
\delta^{1+1} {S}^t & = & 
\frac{1}{2 \Gamma^2} \left [ \frac{1}{X^2 {\alpha}^2} \left \{ 
\Gamma^2 + I'^2 ({\alpha}^2 - X^2 (\Omega^2 - \Omega_{\rm F}^2)) \right \} (\partial_T \psi)^2
+ \gamma_{\rm F}^2 \left \{ {\alpha}^2 - X^2 (\Omega^2 - \Omega_{\rm F}^2) \right \} 
(\partial_x \psi)^2 \right . \nonumber \\
& &  \left . + 2 \frac{I' \gamma_{\rm F}}{{\alpha}^2} 
\{ \Gamma (\Omega - \Omega_{\rm F}) + \Omega_{\rm F}
({\alpha}^2 + X^2 (\Omega - \Omega_{\rm F})^2 ) \} \partial_T \psi \partial_x \psi  \right ]
 , \\
\delta^{2+0} {S}^t & = & 
\frac{\gamma_{\rm F} I'}{- \Gamma} \partial_x \chi - \frac{1}{{\alpha}^2 \Gamma}
(\Omega_{\rm F} \Gamma - I'^2 W) \partial_T \chi.
\end{eqnarray}

The energy flux density in the BTZ coordinates is calculated by
\begin{eqnarray}
{S}^r & = & - T^r_t 
= - \frac{\partial r}{\partial x^{\underline{\lambda}}} \frac{\partial x^{\underline{\kappa}}}{\partial t}
T^{\underline{\lambda}}_{\underline{\kappa}}
= - \left [ \frac{\partial r}{\partial X} \frac{\partial T}{\partial t} T^X_T
+ \frac{\partial r}{\partial X} \frac{\partial \rho}{\partial t} T^X_\rho
\right ] \nonumber \\
& = &  - T^X_T + \Omega_{\rm F} T^X_\rho = \underline{S}^X + \Omega_{\rm F} \underline{M}^X.
\end{eqnarray}
Then, we have the energy flux density of the equilibrium, first and
second-order perturbations of the linear Alfv\'{e}n wave,
\begin{eqnarray}
{\bar{S}}^r &=& \bar{\underline{S}}^X + \Omega_{\rm F} \bar{\underline{M}}^X
= \Omega_{\rm F} \frac{I'}{X} ,\\
\delta {S}^r &=& \delta {\underline{S}}^X + \Omega_{\rm F} \delta \underline{M}^X = 0, \\
\delta^2 {S}^r &=& \delta^2 \underline{S}^X + \Omega_{\rm F} \delta^2 \underline{M}^X
= \delta^{1+1} {S}^r +  \delta^{2+0} {S}^r  ,
\end{eqnarray}
where
\begin{eqnarray}
\delta^{1+1} {S}^r & = & \frac{\Gamma}{X^2}  \partial_T \psi \partial_X \psi
- \Omega_{\rm F} W^{X\lambda T\rho} \partial_\lambda \psi \partial_T \psi
\nonumber \\
& = & - \frac{1}{X^2} \partial_T \psi \left [ 
\left (X^2 - M + \frac{Ma}{2} \Omega_{\rm F} \right ) \partial_X \psi
- \frac{I' X \Omega_{\rm F}}{\Gamma} \partial_T \psi  \right ] , \\
\delta^{2+0} {S}^r & = & -g^{X \rho} \partial_T \chi 
+ \Omega_{\rm F} g^{X \beta} \partial_\beta \chi \nonumber \\
& = & \frac{I'}{X \Gamma} ( - \Gamma + X^2 W \Omega_{\rm F} ) \partial_T \chi
+ \Omega_{\rm F} \underline{\alpha}^2 \partial_X \chi .
\end{eqnarray}
When we use the tortoise coordinate $x$, we have distinct expressions,
\begin{eqnarray}
\delta^{1+1} {S}^r & = & - \frac{\gamma_{\rm F}}{X \Gamma} 
\partial_T \psi \left [ 
\left (X^2 - M + \frac{Ma}{2} \Omega_{\rm F} \right ) \partial_x \psi
+ \frac{I' \Omega_{\rm F}}{\gamma_{\rm F}} \partial_T \psi  \right ] , \\
\delta^{2+0} {S}^r & = & 
\frac{I'}{X \Gamma} [
{\alpha}^2 - X^2 (\Omega - \Omega_{\rm F})(\Omega - 2 \Omega_{\rm F}) ]  \partial_T \chi
- {\alpha}^2 \frac{\Omega_{\rm F} X}{\Gamma} \gamma_{\rm F} \partial_x \chi.
\end{eqnarray}

\section{Energy transport in the BTZ spacetime
\label{appendetbtz}}
We recovered the force-free condition in the BTZ coordinates with
the additional variable $\chi$ so that we have the
energy and momentum conservation,
$\nabla_{{\nu}} T^{{\mu}{\nu}} = 0$
up to the second order of the perturbations.
When we use the time-like Killing vector $\xi_{(t)}^{{\mu}} = (1, 0, 0, 0)$,
we have the energy conservation law in the BTZ coordinates,
\begin{equation}
\nabla_{{\nu}} {S}^{{\nu}}
= \frac{1}{\sqrt{-g}} \partial_{{\nu}} ( \sqrt{-g} {S}^{{\nu}})
= \frac{1}{\sqrt{-g}} \partial_t ( \sqrt{-g} {S}^t)
+ \frac{1}{\sqrt{-g}} \partial_{{i}} ( \sqrt{-g} {S}^{{i}}) = 0 ,
\end{equation}
where 
${S}^{{\nu}} = - \xi_{{\mu}}^{(t)} T^{{\mu}{\nu}}$
is the 4-energy flux density in the BTZ coordinates.
In the case of axisymmetry and translation symmetry with respect to the $z$-direction,
we have
\begin{equation}
\frac{\partial}{\partial t} ( \sqrt{-g} {S}^t)
+ \frac{\partial}{\partial r} ( \sqrt{-g} {S}^r) = 0 .
\end{equation}
It reads,
\begin{align}
& \int_{t_1}^{t_2} dt \int_{r_1}^{r_2 } \frac{\partial}{\partial t} ( \sqrt{-g} {S}^t)
+ \int_{t_1}^{t_2} dt \int_{r_1}^{r_2 } \frac{\partial}{\partial r} ( \sqrt{-g} {S}^r)
\nonumber \\
& =   \int_{r_1}^{r_2 } \sqrt{-g} {S}^t(r,t_2)
-  \int_{r_1}^{r_2 } \sqrt{-g} {S}^t(r,t_1)
+ \int_{t_1}^{t_2} dt \sqrt{-g} {S}^r(r_2,t) 
- \int_{t_1}^{t_2} dt \sqrt{-g} {S}^r(r_1,t) = 0.
\end{align}
When we define the total energy between $r=r_1$ and $r=r_2$ and
the energy flux at $r=r_b$ ($b=1, 2$) by
\begin{eqnarray}
E(t) & = & \int_{r_1}^{r_2} \sqrt{-g} {S}^t(r, t) dr, 
\label{eqofet} \\
F_b (t) & = & \int_{t_1}^{t} \sqrt{-g} {S}^t(r_b, t') dt',
\label{eqoffb}
\end{eqnarray}
respectively, we obtain the conservation quantity as
\begin{equation}
E(t_2) - F_1 (t_2) + F_2 (t_2) = E(t_1).
\end{equation}
When we consider the second order of the perturbations, we have
the quantity with respect to the  energy conservation of the wave,
\begin{equation}
\delta^2 E(t) - \delta^2 F_1 (t) + \delta^2 F_2 (t) = \delta^2 E(t_1),
\end{equation}
where $\displaystyle \delta^2 E(t) = \int_{r_1}^{r_2} \sqrt{-g} \delta^2 {S}^t(r, t) dr$ and
$\displaystyle \delta^2 F_b (t) = \int_{t_1}^{t} \sqrt{-g} \delta^2 {S}^r(r_b, t') dt'$
($b=1, 2$).

\section{Numerical method of 1-D wave equation}
\label{appendnum}

Equations (\ref{newtontoyeq}) with the additional term $-\kappa(x) \psi$ 
in its right-hand side and (\ref{chieq}) are written 
by multi-dimensional time-development equations as,
\begin{eqnarray}
\frac{\partial \psi}{\partial T} &=& 
- \frac{1}{\lambda(x)} \frac{\partial v}{\partial x} -\kappa(x) w, \nonumber \\
\frac{\partial v}{\partial T} &=& - \frac{\partial \psi}{\partial x} , \label{multdimeq} \\
\frac{\partial w}{\partial T} &=& \psi, \nonumber \\
\frac{\partial \chi}{\partial T} &=& - h(x) \frac{\partial}{\partial x}
(b(x) \chi + c(x) u) + k(x) \chi + G, \nonumber \\
\frac{\partial u}{\partial T} &=& - \frac{\partial \chi}{\partial x}, \nonumber \\
\frac{\partial G}{\partial T} &=& g(x), \nonumber
\end{eqnarray}
where $u(x,T)$, $v(x,T)$, $w(x,T)$, and $G(x,T)$ are new variables,
and $\displaystyle h(x)= - \frac{f}{\sqrt{-g} g^{TT}}$, $b(x)=- 2 \sqrt{-g} g^{TX}$, 
$\displaystyle k(x)=\frac{1}{\sqrt{-g} g^{TT}} \frac{\partial}{\partial X} (\sqrt{-g} g^{TX})$, 
$\displaystyle c(x)=\sqrt{-g} g^{XX} f$, $\displaystyle g(x) = \frac{1}{g^{TT}} s$.

We use the two-step Lax-Wendroff scheme for the multi-dimension
time-development equation
\begin{equation}
\frac{\partial \VEC{u}}{\partial t} =
- \VEC{h} \odot \frac{\partial \VEC{w}}{\partial t} + \VEC{f},
\end{equation}
where $\VEC{u}$ is the array of the conserved quantity density, 
$\VEC{w}$ is the array of the flux density of the conserved quantity, 
$\VEC{f}$ is the array of the source density of the conserved variable:
\begin{eqnarray}
\VEC{u}_j^{\overline{n+1}} &=& \VEC{u}_j^n - \frac{\Delta t}{\Delta x} \VEC{h}_j^n \odot
(\VEC{w}_{j+1}^n - \VEC{w}_j^n) + \Delta t \VEC{f}_j^n ,\\
\VEC{u}_j^{n+1} &=& \frac{1}{2} \left [ \VEC{u}_j^n  \VEC{u}_j^{\overline{n+1}}
- \frac{\Delta t}{\Delta x} \VEC{h}_j^n \odot (\VEC{w}_j^{\overline{n+1}} - \VEC{w}_{j-1}^{\overline{n+1}}) 
+ \Delta t \VEC{f}_j^{\overline{n+1}} \right ] .
\end{eqnarray}
Here, we used $\odot$ to express the product of two vectors 
$\VEC{a}=(a_1, a_2, \cdots)^{\rm T}$ and
$\VEC{b}=(b_1, b_2, \cdots)^{\rm T}$,
\begin{equation}
\VEC{a} \odot \VEC{b} \equiv 
\left ( \begin{array}{c} a_1 b_1 \\ a_2 b_2 \\ \vdots \end{array} \right ).
\end{equation}

Equation (\ref{multdimeq}) are given by
\begin{equation}
\VEC{u} = \left ( \begin{array}{c} \psi \\ v \\ w \\ 
\chi \\ u \\ G \end{array} \right ),
\VEC{h} = \left ( \begin{array}{c} \frac{1}{\lambda(x)} \\ 1 \\ 0 \\
h \\ 1 \\ 1 \end{array} \right ),
\VEC{w} = \left ( \begin{array}{c} v \\ \psi \\ 0 \\
b \chi + c u \\ \chi \\ 0 \end{array} \right ),
\VEC{f} = \left ( \begin{array}{c} - \kappa(x) w \\ 0 \\ \psi \\
k \chi + G \\ 0 \\ g  \end{array} \right ).
\end{equation}


\begin{thebibliography}{}
%% \bibitem[EHT Collaboration(2019a)]{eht19a}
%% EHT Collaboration et al. 2019a, \apjl, 875, L1.
%% \bibitem[EHT Collaboration(2019b)]{eht19b}
%% EHT Collaboration et al. 2019b, \apjl, 875, L5.
%% \bibitem[Inda-Koide, Koide, \& Morino(2019)]{indako19}
%% Inda-Koide, M., Koide, S., \& Morino, R. 2019, \apj, 883, 69.
\bibitem[Antolin et al.(2008)]{antolin08}
Antolin, P., Shibata, K., Kudoh, T., Shiota, D., Brooks, D. 2008,
\apj, {\bf 688}, 669.
\bibitem[Banados et al.(1992)]{banados92}
Banados, M., Teitelboim, C., \& Zanelli, J. 1992,
\prl, {\bf 69}, 1849.
%% \bibitem[Abbott et al.(2016)]{abbott16}
%% Abbott, B.P., Abbott, R., Abbott, T. D. 2016, \prl, 116, 241103.
\bibitem[Bellman(2006)]{bellman06}
Bellman, P. M. 2006, {\it Fundamentals of Plasma Physics}
(Cambridge University Press, Cambridge)
%% \bibitem[Biretta(1999)]{biretta99} Biretta, J. A., Sparks, W. B., 
%% Macchetto, F. 1999, \apj, {\bf 520}, 621
\bibitem[Blandford \& Znajek(1977)]{blandford77} Blandford, R. D. \& Znajek, R. 1977,
\mnras, {\bf 179}, 433
%% \bibitem[Davis(1984)]{davis84} Davis, S. F. 1984, NASA Contractor Rep. 172373
%% (ICASE Rep. 84-20) (NASA, Washington).
%% \bibitem[Del Zanna(2007)]{delzanna07}
%% Del Zanna, L., Zanotti, O., Bucciantini, N., \& Londrillo, P. 2007,
%% \aap, 473, 11.
%% \bibitem[Event Horizon Telescope Collaboration (2019a)]{eht19a}
%% Event Horizon Telescope Collaboration 2019, \apjl, 875, L1.
\bibitem[Event Horizon Telescope Collaboration (2019)]{eht19}
Event Horizon Telescope Collaboration 2019, \apjl, 875, L5.
%% \bibitem[Goldreich(1969)]{goldreich69} Goldreich, P. \& Julian, W. H. 1969,
%% \apj, 157, 869.
\bibitem[Gammie et al.(2003)]{gammie03}
Gammie, C. F., McKinney, J. C., \& Toth, G. 2003, \apj, {\bf 589}, 444.
%%% \bibitem[Imamura \& Koide(2019)]{imamura19} 
%%% Imamura, T. \& Koide, S. 2019, \apj, in press.
\bibitem[Jacobseon \& Rodriguez(2019)]{jacobsen19}
Jacobson, T. and Rodriguez, M. J. 2019, \prd, {\bf 99}, 124013.
\bibitem[Kathirgamaraju et al.(2019)]{kathirgamaraju19}
Kathirgamaraju, A., Tchekhovskoy, A., Giannios, D., \& Duran, R. B. 2019,
\mnras, {\bf 484}, L98.
%% \bibitem[Koide(1998)]{koide98} 
%% Koide, S., Shibata, K., and Kudoh, T. 1998
%% \apj, {\bf 495}, L63.
%% \bibitem[Koide(1999)]{koide99} 
%% Koide, S., Shibata, K., and Kudoh, T. 1999,
%% \apj, {\bf 522}, 727.
%%% \bibitem[Koide(2000)]{koide00}  
%%% Koide, S., Meier, D. L., Shibata, K., and Kudoh, T. 2000,
%%% \apj, {\bf 536}, 668.
%% \bibitem[Koide et al.(2002)]{koide02} 
%% Koide, S., Shibata, K., Kudoh, T., \& Meier, D. L. 2002, Science, {\bf 295}, 1688. 
\bibitem[Koide(2003)]{koide03} 
Koide, S. 2003, \prd, {\bf 67}, 104010.
\bibitem[Koide(2004)]{koide04} 
Koide, S. 2004, \apjl, {\bf 606}, L45.
%% \bibitem[Koide \& Baba(2014)]{koide14} 
%% Koide, S. \& Baba, T.  2014, \apj, {\bf 792}, 88
%%% \bibitem[Koide \& Imamura(2018)]{koide18} 
%%% Koide, S. \& Imamura, T.  2018, \apj, {\bf 864}, 173
\bibitem[Koide et al.(2006)]{koide06} 
Koide, S. , Kudoh, T., Shibata, K. 2006, \prd, {\bf 74}, 044005
%%% \bibitem[Komissarov(2001)]{komissarov01} Komissarov, S. S. 2001,
%%% \mnras, {\bf 326}, L41
%%% \bibitem[Komissarov(2002)]{komissarov02} Komissarov, S. S. 2002,
%%% \mnras, {\bf 336}, 759
\bibitem[Komissarov(2004)]{komissarov04} Komissarov, S. S. 2004, \mnras,
{\bf 350}, 1431.
\bibitem[Komissarov(2005)]{komissarov05} 
Komissarov, S. S. 2005, \mnras, {\bf 359}, 801.
%% % \bibitem[Komissarov(2009)]{komissarov09} Komissarov, S. S. 2009, 
%% % Journal of the Korean Phys. Soc., {\bf 54}, 2503
%% \bibitem[Kulkarni(1999)]{kulkarni99} 
%% Kulkarni, S. R. 1999, \nat, {\bf 398}, 389
%% %% \bibitem[Lasota et al. (2017)]{lasota14}
%% %% Lasota, J.-P., Gourgoulhon, E., Abramowicz, M., Tchekhovskoy, A., Narayan, R. 
%% %% 2014, \prd, {\bf 89}, 024041
%% \bibitem[LIGO Scientific Collaboration et al(2017)]{ligo17}
%% LIGO Scientific Collaboration, Virgo Collaboration, Fermi Gamma-ray Burst
%% Monitor, and INTEGRAL 2017, \apjl, {\bf 848}, L13.
\bibitem[McKinney(2006)]{mckinney06} 
McKinney, J. C. 2006, \mnras, {\bf 368}, 1561
\bibitem[McKinney(2009)]{mckinney09}
McKinney, J. C. \& Blandford, R. D. 2009,
\mnras, {\bf 394}, L126
%%% \bibitem[McKinney et al.(2013)]{mckinney13}
%%% McKinney, J. C., Tchekhovskoy, A., Blandford, R. D. 2013,
%%% \mnras, {\bf 423}, 3083
%% \bibitem[Menon \& Dermer(2005)]{menon05}
%% Menon, G. \& Dermer, C. D. 2005, \apj, {\bf 635}, 1197.
%% \bibitem[Mirabel(1994)]{mirabel94} 
%% Mirabel, I. F., \& Rodriguez, L. F. 1994,
%% \nat, {\bf 374}, 141
\bibitem[Mizuno et al. (2004)]{mizuno04}
Mizuno, Y., Yamada, S., Koide, S., \& Shibata, K. 2004,
\apj, {\bf 615}, 389.
\bibitem[Musielak(2007)]{musielak07}
Musielak, Z. E., Routh, S. \& Hammer, R. 2007,
\apj, {\bf 659}, 650.
\bibitem[Nagataki(2009)]{nagataki09}
Nagataki, S. 2009, \apj, {\bf 704}, 937.
\bibitem[Noda et al.(2020)]{noda20}
Noda, S., Nambu, Y., Tsukamoto, T., Takahashi, M., 2020, \prd, 101, 023003.
%% %% \bibitem[Palenzuela(2010)]{palenzuela10}
%% %% Palenzuela, C., Garrett, T.,Lehner, L., Liebling, S. L. 2010,
%% %% \prd, {\bf 82}, id.044045
\bibitem[Paschalidis et al.(2015)]{paschalidis15}
Paschalidis, V., Ruiz, M., \& Shapiro, S. L. 2015,
\apj, {\bf 806}, L14.
%% %% \bibitem[Pearson(1987)]{pearson87} 
%% %% Pearson, T. J. \& Zensus, J. A. 1987, in {\it Superluminal Radio Sources},
%% %% edited by J. A. Zensus and T. J. Pearson 
%% %% (Cambridge Univ., London), p. 1.
%%% \bibitem[Penrose(1969)]{penrose69}
%%% Penrose, R. 1969, Nuovo Cimento, {\bf 1}, 252
\bibitem[Porth et al.(2019)]{porth19}
Porth, O., Chatterjee, K., Narayan, R. et al. 2019, \apjs, {\bf 243}, id. 26.
%% % \bibitem[Punsly(1989)]{punsly89}
%% % Punsly, B. \& Coroniti, F. V. 1989, \prd, {\bf 40}, 3834
%% \bibitem[Radice \& Rezzolla(2013)]{radice13}
%% Radice, D \& Rezzolla, L. 2013, in Astronomical Society of the Pacific
%% Conference Series, Vol. 474, Numerical Modeling of Space Plasma Flows
%% (ASTRONUM2012), ed. N. V. Pogorelov, E. Audit, \& G. P. Zank, 25.
%%% \bibitem[Tchekhovskoy(2010)]{tchekhovskoy10}
%%% Tchekhovskoy, A., Narayan, R., McKinney, J. C. 2010, \apj, {\bf 711}, 50
%%% \bibitem[Tingay(1995)]{tingay95} 
%%% Tingay. S. J., et al. 1995, \nat, {\bf 374}, 141
%%% \bibitem[Toma \& Takahara(2016)]{toma16}
%%% Toma, K. \& Takahara, F. 2016,
%%% PTEP, {\bf 2016}, 063E01
%%% \bibitem[Wagh(1989)]{wagh89}
%%% Wagh, S. M., Dadhich, N., Physics Reports, {\bf 183}, 139 (1989).
%% % \bibitem[Takahashi(1990)]{takahashi90} M. Takahashi, S. Nitta,
%% % Y. Tatematsu, and A. Tomimatsu, Astrophys. J. {\bf 363}, 206 (1990).
%% % \bibitem[Hirotani(1992)]{hirotani92} K. Hirotani, M. Takahashi,
%% % S.-Y. Nitta, and A. Tomimatsu, Astrophys. J. {\bf 386}, 455 (1992).
%% % \bibitem[Press(1972)]{press72}
%% % W. H. Press, S. A. Teukolsky, Nature, {\bf 238}, 211 (1972).
%% % \bibitem[Lightman(1975)]{lightman75}
%% % A. P. Lightman, W. H. Press, R. H. Price, \& S. A. Teukolsky,
%% % {\it Problem book in relativity and gravitation},
%% % (Princeton Univ. Press, Princeton, 1975).
%% % % \bibitem[Koide(2008)]{koide08} 
%% % % S. Koide, Physical Review D, {\bf 78}, 125026 (2008).
%%% 13
\bibitem[Ruiz et al.(2016)]{ruiz16}
Ruiz, M., Lang, R. N., Paschalidis, V., \& Shapiro, S. L. 2016,
\apj, {\bf 824}, L6.
%%%%
%% % \bibitem[Weinberg(1972)]{weinberg72} 
%% % S. Weinberg, {\it Gravitation and Cosmology} 
%% % (John Wiley \& Sons, New York, 1972). 
\bibitem[Uchida(1997a)]{uchida97a}
Uchida, T. 1997a, Monthly Notices of the Royal Astronomical Society, {\bf 286}, 931.
\bibitem[Uchida(1997b)]{uchida97b}
Uchida, T. 1997b, Monthly Notices of the Royal Astronomical Society, {\bf 291}, 125.
%%% \bibitem[Znajek(1977)]{znajek77}
%%% Znajek, R. 1977, Monthly Notices of the Royal Astronomical Society, {\bf 179}, 457.
\end{thebibliography}
\end{document}